\documentclass[onecolumn]{article}
\usepackage{todonotes}
\usepackage{psfrag}
\usepackage{tikz}
\usepackage{svg}
\usepackage{amsmath,amsthm,amssymb}
\usepackage{url}

\usepackage{amsmath,amssymb,amsfonts,amsthm,enumerate,multirow,mathtools}
\usepackage{color}
\usepackage{siunitx}
\usepackage{parskip}
\usepackage[hidelinks]{hyperref}
\usepackage{placeins}
\usepackage{booktabs}
\usepackage{algorithm,algpseudocode}
\usepackage[affil-it]{authblk}
\usepackage[margin=2.5cm]{geometry}
\usepackage[round]{natbib}

\usepackage{fancyhdr}
\usepackage{hyperref}
\hypersetup{
    colorlinks=true,
    linkcolor=blue,
    filecolor=magenta,      
    urlcolor=cyan,
    pdftitle={Overleaf Example},
    pdfpagemode=FullScreen,
    }

\DeclareMathOperator*{\argmax}{arg\,max}

\title{Intragranular Strain Estimation in Far-Field Scanning X-ray Diffraction using a Gaussian Processes}

\author[1]{N.A. Henningsson}
\author[2]{J.N. Hendriks}

\affil[1]{Division of Solid Mechanics, Lund University, Lund, Sweden}
\affil[2]{School of Engineering, The University of Newcastle, Callaghan NSW 2308, Australia}

\begin{document}
\newcommand{\coverTitle}{Intragranular Strain Estimation in Far-Field Scanning X-ray Diffraction using a Gaussian Processes}
\newcommand{\coverAuthors}{N.A. Henningsson \& J.N. Hendriks}
\newcommand{\coverStatus}{Preprint}

\pagestyle{fancy}
\rfoot{Page \thepage}
\lfoot{\coverStatus}
\cfoot{\coverAuthors}

\begin{titlepage}
    \begin{center}
        {\large \em \coverStatus}
        
        \vspace*{2.5cm}
        %
        {\Huge \bfseries \coverTitle  \\[0.4cm]}
        %
        {\Large \coverAuthors \\[2cm]}

        \renewcommand\labelitemi{\color{red}\large$\bullet$}
        \begin{itemize}
            \item {\Large \textbf{ This is a preprint version, the final version is published with open access in:
            \newline
            \newline
            Journal of Applied Crystallography 54(4)
            \newline
            DOI: 10.1107/S1600576721005112 
            \newline
            \newline
            Please refer to the final published version for citing. \newline
            url: \url{https://scripts.iucr.org/cgi-bin/paper?S1600576721005112}
            }} \\[0.4cm]
            
        \end{itemize}
        
        \vfill
        
        \vfill
    \end{center}
\end{titlepage}

\maketitle

\begin{abstract}
A new method for estimation of intragranular strain fields in polycrystalline materials based on scanning three-dimensional X-ray diffraction data (scanning-3DXRD) is presented and evaluated. Given an apriori known anisotropic compliance, the regression method enforces the balance of linear and angular momentum in the linear elastic strain field reconstruction. By using a Gaussian Process (GP), the presented method can yield a spatial estimate of the uncertainty of the reconstructed strain field. Furthermore, constraints on spatial smoothness can be optimised with respect to measurements through hyperparameter estimation. These three features address weaknesses discussed for previously existing scanning-3DXRD reconstruction methods and, thus, offers a more robust strain field estimation. The method is twofold validated; firstly by reconstruction from synthetic diffraction data and, secondly, by reconstruction of previously studied tin (Sn) grain embedded in a polycrystalline specimen. Comparison against reconstructions achieved by a recently proposed algebraic inversion technique is also presented. It is found that the GP regression consistently produces reconstructions with lower root mean squared errors, mean absolute errors and maximum absolute errors across all six components of strain.
\end{abstract}

\section{Introduction}
Three-dimensional X-ray Diffraction (3DXRD), as pioneered by \cite{HenningDisertation} and co-workers, is a nondestructive materials probe for the study of bulk polycrystalline materials. The experimental technique is typically implemented at synchrotron facilities where access to hard X-rays (\(\geq\) 10 keV) facilitate the study of dense materials with sample dimensions in the mm range. In contrast to powder diffraction, 3DXRD enables studies on a per grain basis, which requires that the Debye-Scherrer rings consist of a set of well-defined, separable single crystal peaks. To achieve this, the beam and sample dimensions must be selected accordingly, limiting the number of grains illuminated per detector readout. By various computer aided algorithms, c.f \citep{Lauridsen2001}, the single crystal diffraction peaks may be segmented and categorised on a per grain basis, enabling the study of individual crystals within a sample. Typical quantities retrieved from such analysis are the grain average strain and average orientation \cite{Poulsen2001Strain}, \cite{Oddershede2010}. From further analysis it also possible to retrieve an approximate grain topology map, \cite{Poulsen2003FBP}, \cite{Poulsen2003ART}, \cite{Markussen2004}, \cite{Alpers2006}.

Reducing the X-ray beam cross-section to sub grain dimensions not only allows for the study of samples with large amounts of grains, but also enables the investigation of intragranular variations. This special case of 3DXRD is commonly referred to as scanning-3DXRD since, to acquire a full data set, the narrow beam must be scanned across the sample. In this setting, it is possible to measure the diffraction signal from approximate line segments across the grains, collecting information on the intragranular structure. Any inversion procedure, in pursuit of such intragranular quantities, then poses a rich tomography problem where the ray transform typically involves higher-order tensorial fields.

Recent advances in Diffraction Contrast Tomography (DCT) \cite{Reischig2020} show promising results for inversion for both orientation and strain fields in 3D with intragranular resolution. In scanning-3DXRD where higher angular resolution on scattering vectors are achieved at the cost of diffraction peak resolution \cite{Nervo2014}, multiple proposals for inversion operating solly from scattering vectors has been made. Initially, \citep{Hayashi2015} and \citep{Hayashi2017} proposed a method for a per voxel strain refinement to approximate intragranular strains using scanning-3DXRD data. Unfortunately, this procedure was shown to introduce bias in the reconstruction related to strain state \citep{Hayashi2019}, \citep{Hektor2019}. These obstacles were later overcome by \citep{Henningsson2020}, who proposed an inversion method that takes the tomographic nature of the problem into account. As has been pointed out by several other authors, c.f \citep{Margulies2002} and \citep{Lionheart2015}, the sampling of strain is not uniform in 3DXRD and, as a result, some additional constraints on the reconstructed field is often desirable. \citep{Henningsson2020} proposed a simple smoothing constraint to each of the strain components with success. However, the parameter selection, and the physical interpretation of these constraints, are not well defined. 

For powder diffraction type data excellent progress to overcome the weaknesses highlighted above has been made using a so-called Gaussian Process \cite{hendriks2019}. In this current work, we adapt the Gaussian Process (GP) framework to scanning-3DXRD and extend it to a wider class of anisotropic materials. This framework allows for the introduction of a static equilibrium constraint, which ensures that the retrieved strain reconstruction will satisfy the balance of both angular and linear momentum. The GP naturally incorporates spatial correlation in the predicted fields via a covariance function, which, together with the equilibrium prior, replaces the need for previously used smoothing constraints. Moreover, the GP produces an estimate of the uncertainty in the reconstructed strain field, as a biproduct of regression. In total, the presented regression procedure addresses several weaknesses of previous work and provides a tool for uncertainty estimation in the reconstructed strain fields.

In the following, we describe how the GP regression for scanning-3DXRD can be performed and validated. The paper is structured as follows. In section \ref{sec:Diffraction Measurements}, we outline the experimental setup used in scanning-3DXRD together with the associated commonly used mathematical framework. We further briefly describe the necessary data analysis routinely performed in 3DXRD to achieve grain maps. In section \ref{sec:Measurement Model}, we describe the mathematical measurement model and derive the necessary transformations from diffraction data to strain related quantities. Section \ref{sec:Regression Procedure} is dedicated to deriving the Gaussian Process framework. Validation of the method is presented in section \ref{sec:Validation} based on reconstructions both from simulated data as well as a previously investigated columnar tin (Sn) grain. Finally, in sections \ref{sec:Discussion}-\ref{sec:Conclusions}, we conclude and discuss our findings and provide a brief outlook.

\section{Diffraction Measurements}\label{sec:Diffraction Measurements}

\subsection{Experimental Acquisition}\label{sec:Experimental Acquisition}

\begin{figure}[H]
    \centering
    \includegraphics[scale=0.6]{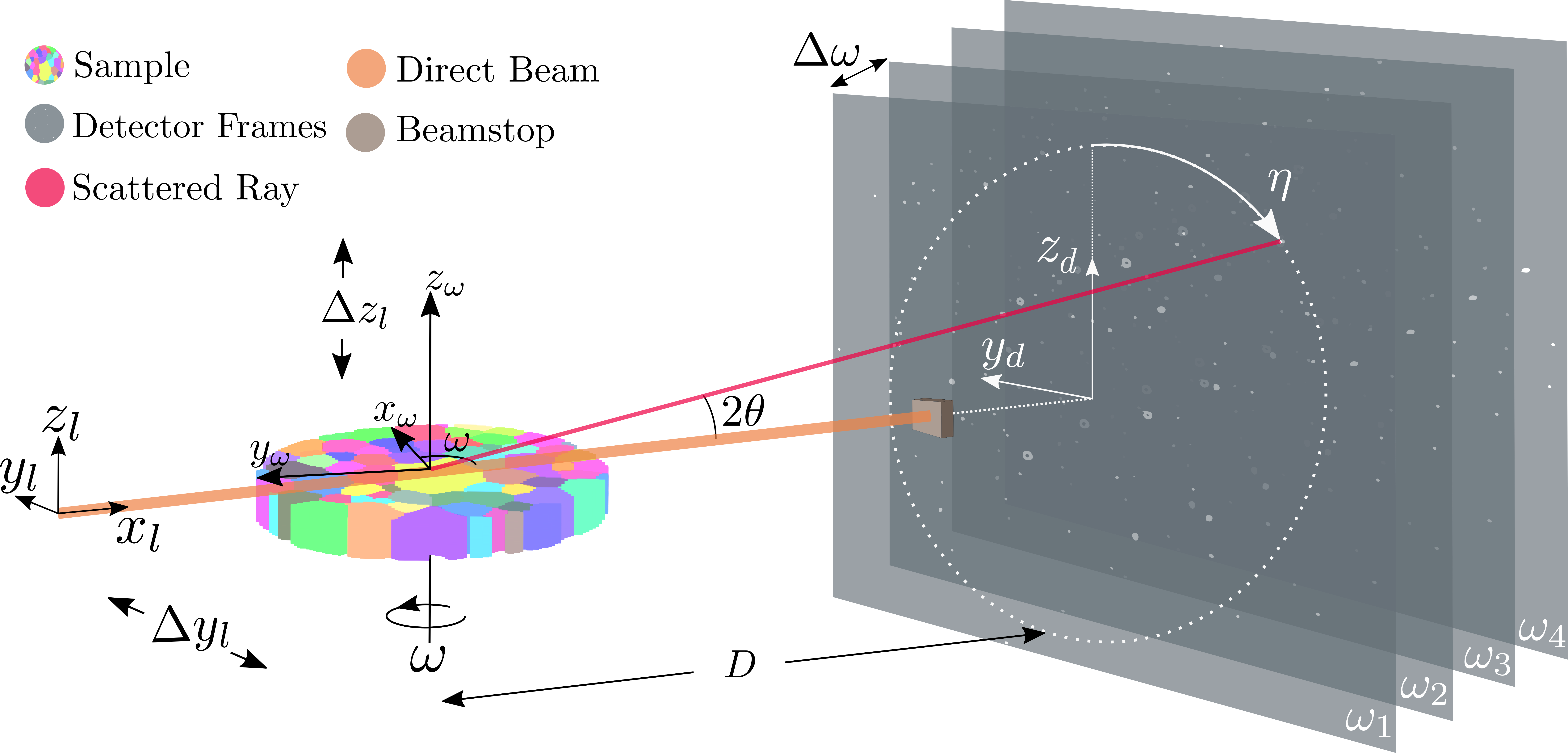}
    \caption{Scanning-3DXRD experimental setup. The sample coordinate system (subscript \(\omega\)) is attached to the sample turntable while the laboratory (subscript \(l\)) coordinate system is fixed in relation to the sample. The sample is rotated and translated in the \(y_l\)-\(z_l\)-plane across the beam to record diffraction from the full volume. (modified from \citep{Henningsson2020})}
    \label{fig:experimental_setup}
\end{figure}

In scanning-3DXRD, a polycrystalline specimen is placed on a sample stage associated with an attached coordinate system (\(\hat{\boldsymbol{x}}_{\omega}\),\(\hat{\boldsymbol{y}}_{\omega}\),\(\hat{\boldsymbol{z}}_{\omega}\)). The sample stage commonly has several degrees of freedom, some of which are used for initial alignment and calibration and others for data collection. Since the calibration procedure is the same for all 3DXRD type of experiments, here we only describe the degrees of freedom related to data acquisition; for details on calibration see \citep{Oddershede2010} and \citep{Edmiston2011}. A fixed laboratory coordinate system (\(\hat{\boldsymbol{x}}_{l}\),\(\hat{\boldsymbol{y}}_{l}\),\(\hat{\boldsymbol{z}}_{l}\)) is introduced and related to the sample coordinate system through a positive rotation about \(\hat{\boldsymbol{z}}_{l}\) and a translation in the \(\hat{\boldsymbol{y}}_{l}\)-\(\hat{\boldsymbol{z}}_{l}\)-plane (Figure \ref{fig:experimental_setup}). For a given sample position (\(y_{l}\),\(z_{l}\)), rotation in \(\omega\) is performed in discrete steps of \(\Delta \omega\). The scattered intensity in each \(\Delta \omega\) rotation interval is generally integrated during the acquisition, resulting in a series of frames for each (\(y_l,z_l\)) position. After any necessary spatial distortion corrections have been made, the raw pixelated image stacks (\(y_d,z_d,\omega\)) can be segmented into separate connected regions of diffracted intensity for which centroids and average intensities can be calculated. The resulting data set is 6D, with each diffraction peak average intensity and detector centroid (\(\theta,\eta\)) mapping to a sample stage setting (\(y_l,z_l,\omega\)).

\subsection{Laue Equations and scattering notation}\label{subsec:Laue Equations and scattering notation}

From the diffraction peak centroids (\(\theta,\eta\)) it is possible to compute scattering vectors, \(\boldsymbol{G}\), defined in laboratory frame as
\begin{equation}
    \boldsymbol{G}_l = \dfrac{2\pi}{\lambda}
    \begin{bmatrix}
    \cos(2\theta)-1\\
    -\sin(2\theta)\sin(\eta)\\
    \sin(2\theta)\cos(\eta)
    \end{bmatrix}.
    \label{eq:scattering_vector_defenition}
\end{equation}
Using the notation of \cite{HenningDisertation} and considering that the Laue equations are fulfilled during diffraction, we may also express the scattering vectors as
\begin{equation}
    \boldsymbol{G}_l=\boldsymbol{\Omega}\boldsymbol{U}\boldsymbol{B}\boldsymbol{G}_{hkl},
    \label{eq:Gl<=Ghkl}
\end{equation}
where \(\boldsymbol{\boldsymbol{\Omega}}\) and \(\boldsymbol{\boldsymbol{U}}\) are unitary square 3x3 rotation matrices describing, respectively, the turntable rotation around \(\hat{\boldsymbol{z}}_{\omega}\) and the crystal unit cell orientation with respect to the \(\omega\) coordinate system. The matrices \(\boldsymbol{U}\) and \(\boldsymbol{B}\) can now be uniquely defined as the polar decomposition of their inverse product, \((\boldsymbol{U}\boldsymbol{B})^{-1}\), in which the rows contain the real space unit cell lattice vectors \(\boldsymbol{a},\boldsymbol{b}\) and \(\boldsymbol{c}\) described in the sample \(\omega\)-coordinate system as
\begin{equation}
    (\boldsymbol{U}\boldsymbol{B})^{-1} =\begin{bmatrix}\boldsymbol{a}^T\\\boldsymbol{b}^T\\\boldsymbol{c}^T\end{bmatrix}=
    \begin{bmatrix}
    a_1 & a_2 & a_3 \\
    b_1 & b_2 & b_3 \\
    c_1 & c_2 & c_3 \\
    \end{bmatrix}.
    \label{eq:def_UBI}
\end{equation}
The integer vector \(\boldsymbol{G}_{hkl}=\begin{bmatrix}h&k&l\end{bmatrix}^T\) holds the Miller indices.

\subsection{Grain Mapping}\label{subsec:Grain Mapping}
\begin{figure}[H]
    \centering
    \includegraphics[scale=0.75]{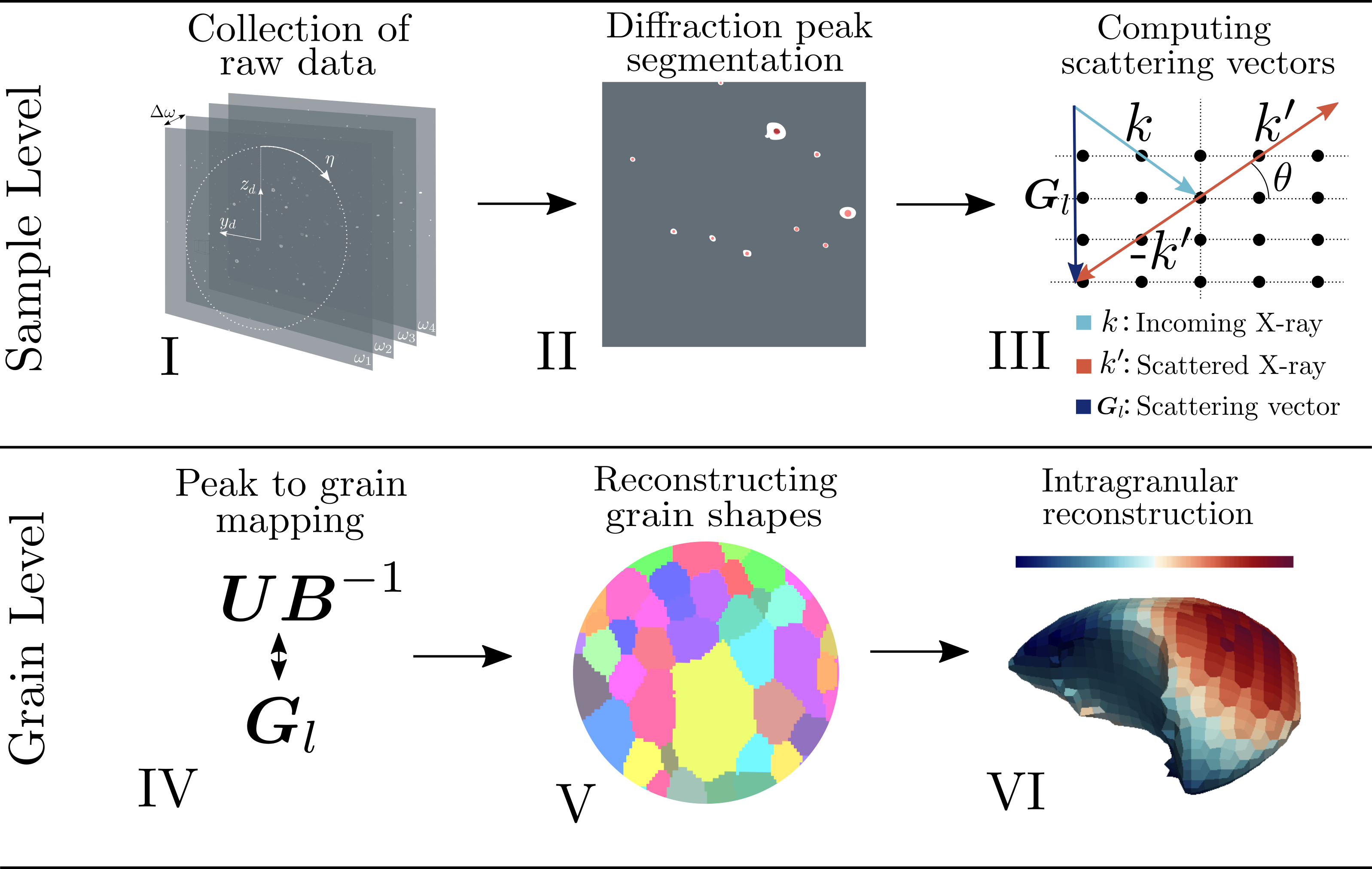}
    \caption{Simplified schematic of analysis steps commonly performed on scanning-3DXRD data. From raw detector data (I) the per peak centroids \(\eta,\theta\) and average intensities are retrieved (II). The scattering vectors can then be computed (III) and inputted to a peak-grain mapping algorithm (IV). From the produced maps per grain shape reconstruction can take place (V). Finally, intragranular quantities may be sought (VI). }
    \label{fig:3dxrd_analysis_schematic}
\end{figure}

Given a measured set of scattering vectors, the procedure known as grain mapping is concerned with finding a set of uniform crystals that explain the data. In this setting, grains are represented by their average \( (\boldsymbol{U}\boldsymbol{B})^{-1}\) matrices together with their real space centroid coordinates. To contextualise the grain mapping procedure a simplified schematic of the scanning-3DXRD analysis steps are presented in Figure \ref{fig:3dxrd_analysis_schematic}.
In essence, the grain mapping procedure results in a map between diffraction peaks and individual average grain \( (\boldsymbol{U}\boldsymbol{B})^{-1}\) matrices and centroids. The diffraction peaks associated with a single grain can be extracted from such peak-grain maps and grain shape reconstruction can be performed by tomographic methods, c.f \citep{Poulsen2003FBP} and \citep{Alpers2006}, utilising the scattered intensity associated to each diffraction peak. The peak-grain maps also enable studies on a per-grain basis, something which simplifies analysis both conceptually as well as computationally. Software for performing grain mapping is freely available in the ImageD11 package \cite{ID11} and further algorithm details can be found in \citep{Oddershede2010} and \citep{Edmiston2011}. In this paper we are concerned with reconstruction of intragranular strain, and thus we focus on the the final step of analysis, and proceed with the assumption that all preceding quantities have been computed for.

\section{Measurement Model}\label{sec:Measurement Model}

\subsection{Strain revealing transformations}\label{subsec:From diffraction peaks to directional strains}
\citep{Henningsson2020} described the procedure to calculate strains in individual lattice planes from scanning-3DXRD measurements via the Bragg equations as first laid out by \citep{Poulsen2001Strain} and \citep{Margulies2002}. To enrich the framework, allow for consistent use of the Laue equations and make it clear how the integration of strain can take place, here we adopt a different route, rewriting the Laue equations and performing a first-order Taylor series expansion. We start by recollecting that the 3x3 continuum deformation gradient tensor, \(\boldsymbol{F}\), should have the property that
\begin{equation}
    \boldsymbol{v} = \boldsymbol{F}\boldsymbol{v}_0,
    \label{eq:def_F}
\end{equation}
where \(\boldsymbol{v}_0\) is a vector in the reference configuration and \(\boldsymbol{v}\) is the corresponding deformed vector. Applying this to a crystal reference unit cell (\(\boldsymbol{a}_0,\boldsymbol{b}_0,\boldsymbol{c}_0\)) given in sample \(\omega\)-coordinate system and collecting the equation in matrix format, we find that
\begin{equation}
    \begin{bmatrix}\boldsymbol{a}&\boldsymbol{b}&\boldsymbol{c}\end{bmatrix}= \boldsymbol{F}\begin{bmatrix}\boldsymbol{a}_0&\boldsymbol{b}_0&\boldsymbol{c}_0\end{bmatrix}.
    \label{eq:F_on_cell}
\end{equation}
With \eqref{eq:def_UBI} this allows us to identify that
\begin{equation}
    \boldsymbol{F} = (\boldsymbol{U}\boldsymbol{B})^{-T}(\boldsymbol{U}_0\boldsymbol{B}_0)^{T},
    \label{eq:F_as_Laue}
\end{equation}
where \(\boldsymbol{U}_0\) and \(\boldsymbol{B}_0\) define an undeformed crystal lattice. We can now relate the quantities involved in the Laue equations \eqref{eq:scattering_vector_defenition} to the strain tensor by considering that the infinitesimal strain tensor in sample \(\omega\)-coordinate system is defined as
\begin{equation}
    \boldsymbol{\epsilon}_{\omega} = \dfrac{1}{2}( \boldsymbol{F}^T +  \boldsymbol{F}) - \boldsymbol{I},
    \label{eq:def_strain}
\end{equation}
where \(\boldsymbol{I}\) is the identity tensor. Insertion of \eqref{eq:F_as_Laue} into \eqref{eq:def_strain} gives
\begin{equation}
    \boldsymbol{\epsilon}_{\omega} = \dfrac{1}{2}\bigg( (\boldsymbol{U}_0\boldsymbol{B}_0)(\boldsymbol{U}\boldsymbol{B})^{-1} +  (\boldsymbol{U}\boldsymbol{B})^{-T}(\boldsymbol{U}_0\boldsymbol{B}_0)^{T}\bigg) - \boldsymbol{I}.
    \label{eq:strain_Laue}
\end{equation}
The observable quantity in 3DXRD are scattering vectors and a useful formulation must therefore relate \(\boldsymbol{\epsilon}_{\omega}\) to \(\boldsymbol{G}_{\omega}\) together with the known quantities \(\boldsymbol{U}_0\) and \(\boldsymbol{B}_0\). To achieve this we consider the strain in a single direction, introducing the unit vector \(\hat{\boldsymbol{\kappa}}\) into \eqref{eq:strain_Laue} as
\begin{equation}
    \hat{\boldsymbol{\kappa}}^T\boldsymbol{\epsilon}_{\omega}\hat{\boldsymbol{\kappa}} = \dfrac{1}{2}\hat{\boldsymbol{\kappa}}^T\bigg( (\boldsymbol{U}_0\boldsymbol{B}_0)(\boldsymbol{U}\boldsymbol{B})^{-1} +  (\boldsymbol{U}\boldsymbol{B})^{-T}(\boldsymbol{U}_0\boldsymbol{B}_0)^{T}\bigg)\hat{\boldsymbol{\kappa}} - 1.
    \label{eq:directional_strain}
\end{equation}
The problem is now to select \(\hat{\boldsymbol{\kappa}}\) such that the right hand side reduces to an observable quantity. One possible selection is to sample the strain parallel to the scattering vector
\begin{equation}
    \hat{\boldsymbol{\kappa}}=\dfrac{\boldsymbol{G}_{\omega}}{|| \boldsymbol{G}_{\omega} ||} = \dfrac{\boldsymbol{U}\boldsymbol{B}\boldsymbol{G}_{hkl}}{|| \boldsymbol{G}_{\omega} ||}.
\end{equation}
Insertion into \eqref{eq:directional_strain} now reduces to 
\begin{equation}
    \hat{\boldsymbol{\kappa}}^T\boldsymbol{\epsilon}_{\omega}\hat{\boldsymbol{\kappa}} =\dfrac{\boldsymbol{G}_{\omega}^T\boldsymbol{G}^{(0)}_{\omega}}{\boldsymbol{G}_{\omega}^T\boldsymbol{G}_{\omega}}  - 1,
    \label{eq:point_directional_strain}
\end{equation}
where, from \eqref{eq:Gl<=Ghkl}, we have
\begin{equation}
\begin{split}
    &\boldsymbol{G}^{(0)}_{\omega} = \boldsymbol{\Omega}^{-1}\boldsymbol{G}^{(0)}_l = \boldsymbol{U}_0\boldsymbol{B}_0\boldsymbol{G}_{hkl}, \\
    &\boldsymbol{G}_{\omega} = \boldsymbol{\Omega}^{-1}\boldsymbol{G}_l = \boldsymbol{U}\boldsymbol{B}\boldsymbol{G}_{hkl}.
\end{split}
\end{equation}
This selection of unit vector \(\hat{\boldsymbol{\kappa}}\) not only guarantees that \(\boldsymbol{\epsilon}_{\omega}\) is the only unknown in \eqref{eq:point_directional_strain}, but further spreads the sampling of strain to all directions defined by the measured set of scattering vectors \(\boldsymbol{G}_{\omega}\). For high X-ray energies, although not uniform, this spread is typically good \cite{Lauridsen2001}, explaining why, in general strain reconstruction is possible in 3DXRD.

\subsection{Tensorial Ray Transform}
So far we have worked with equations \eqref{eq:Gl<=Ghkl}-\eqref{eq:point_directional_strain} as if scattering occurred from a single point. This is typically the approximation made in 3DXRD when only grain average properties are required. For scanning-3DXRD, when pursuing intragranular quantities, we must consider that scattering takes place from grain sub regions, illuminated by the narrow X-ray beam. In fact, if the scattered intensity is the same from all points within the grain, the scattering vectors known to us from the experiment are average quantities over regions, \(\mathcal{R}\), within the grain such that
\begin{equation}
    \langle \boldsymbol{G}_{\omega}\rangle =\dfrac{1}{V}\int_{\mathcal{R}} \boldsymbol{G}_{\omega} dv = \dfrac{1}{V}\int_{\mathcal{R}} \boldsymbol{U}\boldsymbol{B}\boldsymbol{G}_{hkl} dv,
    \label{eq:average_scattering_vector}
\end{equation}
where \(V\) is the total volume of \(\mathcal{R}\), \(dv\) the differential on \(\mathcal{R}\) and \(\langle \cdot \rangle \) indicates volume average. We run now the risk of invalidating our previous results \eqref{eq:point_directional_strain} since the local scattering vectors \(\boldsymbol{G}_{\omega}=\boldsymbol{G}_{\omega}(x_{\omega},y_{\omega},z_{\omega})\) are unknown in scanning-3DXRD. To maintain a useful expression we must further transform \eqref{eq:point_directional_strain} into an equation in \(\langle \boldsymbol{G}_{\omega}\rangle\) rather than \(\boldsymbol{G}_{\omega}\). However, since the strain is nonlinear in \(\boldsymbol{G}_{\omega}\), direct volume integration of \eqref{eq:point_directional_strain} is not possible. Fortunately though, we may obtain an approximation by Taylor expansion of \eqref{eq:point_directional_strain} at \(\boldsymbol{G}_{\omega}=\boldsymbol{G}_{\omega}^{(0)}\) to first order
\begin{equation}
    \hat{\boldsymbol{\kappa}}^T\boldsymbol{\epsilon}_{\omega}\hat{\boldsymbol{\kappa}} \approx 1 - \dfrac{\boldsymbol{G}_{\omega}^T\boldsymbol{G}^{(0)}_{\omega}}{(\boldsymbol{G}^{(0)}_{\omega})^T\boldsymbol{G}^{(0)}_{\omega}}.
    \label{eq:average_point_directional_strain_taylor}
\end{equation}
By selecting a uniform reference configuration in space, integration of \eqref{eq:average_point_directional_strain_taylor} now gives, with \eqref{eq:average_scattering_vector}, that
\begin{equation}
    y=\dfrac{1}{V}\int_{\mathcal{R}}\hat{\boldsymbol{\kappa}}^T\boldsymbol{\epsilon}_{\omega}\hat{\boldsymbol{\kappa}} dv \approx1 - \dfrac{\langle \boldsymbol{G}_{\omega}\rangle^T\boldsymbol{G}^{(0)}_{\omega}}{(\boldsymbol{G}^{(0)}_{\omega})^T\boldsymbol{G}^{(0)}_{\omega}},
    \label{eq:average_point_directional_strain}
\end{equation}
where we introduce the scalar average strain measure \(y=y(\hat{\boldsymbol{\kappa}})\).

Finally, in any inversion scheme where \(\boldsymbol{\epsilon}_{\omega}\) constitute the free variables, we must be able to execute the forward model that is the integral of \eqref{eq:average_point_directional_strain}. For this purpose the direction of strain,  \(\hat{\boldsymbol{\kappa}}\), must be approximated. Using the already introduced assumption that \(\boldsymbol{G}_{\omega}\) varies weakly on \(\mathcal{R}\) we can write
\begin{equation}
    \hat{\boldsymbol{\kappa}} \approx \dfrac{\langle \boldsymbol{G}_{\omega}\rangle}{|| \langle \boldsymbol{G}_{\omega}\rangle ||}.
    \label{eq:kappa_approx}
\end{equation}
We note that, equally, the approximation \(\hat{\boldsymbol{\kappa}} \approx \boldsymbol{G}^{(0)}_{\omega}/|| \boldsymbol{G}^{(0)}_{\omega}||\) could have been made.

In conclusion, \eqref{eq:average_point_directional_strain} and \eqref{eq:kappa_approx} relate the measured average scattering vectors, \(\langle \boldsymbol{G}_{\omega}\rangle\), to the underlying strain field, \(\boldsymbol{\epsilon}_{\omega}(x_{\omega},y_{\omega},z_{\omega})\), with the strain tensor being the only involved unknown quantity.

\subsection{Estimated Uncertainty}
To finalise the measurement model we introduce an additive Gaussian error \(e\) into \eqref{eq:average_point_directional_strain} representing measurement uncertainty. Furthermore, to simplify both computation and further analytical derivations we approximate the volume integral over \(\mathcal{R}\) by a corresponding line integral going through the geometrical centre of this region. In total, we have the measurement model,
\begin{equation}
    y =\dfrac{1}{L}\int_{\mathcal{L}}\hat{\boldsymbol{\kappa}}^T\boldsymbol{\epsilon}_{\omega}\hat{\boldsymbol{\kappa}}dl + e,
    \label{eq:measurement_model}
\end{equation}
where \(L\) is the length of the line segment \(\mathcal{L}\) and \(e\) is the additive normally distributed noise
\begin{equation}
    e\sim\mathcal{N}(\mathbb{E}[e],\mathbb{C}[e,e]),
\end{equation}
 with expectation value \(\mathbb{E}[e]\) and covariance \(\mathbb{C}[e,e]\).

The measurement noise is assumed to be zero mean (\(\mathbb{E}[e_i] = 0\)) and independent (\(\mathbb{C}[e_i,e_j] = 0\)) with a selected variance motivated by previous work, \citep{Borbely2014}, \citep{Henningsson2020},
\begin{equation}
    \mathbb{C}[e_i,e_i] = 
    \Bigg(\dfrac{\partial \varepsilon}{\partial r} \Bigg)^{-2},
    \label{eq:weight}
\end{equation}
where \(r=r(\theta)\) is the radial detector coordinate and the indices \(i\) and \(j\) indicate unique measurements. Other estimations of \(\mathbb{C}[e_i,e_i]\) are possible, importantly, though, the variance should depend on the scattering angle \(2\theta\), as, for a 2D detector with uniform pixel size, the measurement uncertainty increases with decreasing scattering angle. 

\section{Regression Procedure}\label{sec:Regression Procedure}

Equation \eqref{eq:measurement_model} is a ray transform that contains information on the average directional strain for a region within the grain. The problem to reconstruct the full strain tensor field from a series of such measurements is therefore tomographic in nature, and the measurements, \(y\), are highly spatially entangled as the regions \(\mathcal{L}\) in general will intersect. A collection of \(N\) measurements, 
\begin{equation}
    \boldsymbol{y}=\begin{bmatrix}y_1 & y_2 & ... &  y_j & ... y_N \end{bmatrix}^T,
    \label{eq:y_full_measurement_set}
\end{equation}
could represent the second member of a linear equation system where \eqref{eq:measurement_model} is used to form a system matrix and a vector of unknown strains defined on some finite basis. This has been described in \citep{Henningsson2020} for a voxel basis, using a weighted least-squares (WLSQ) approach to retrieve the strain field. As we will discuss in the following section, in this work we adapt these ideas to a Gaussian Process framework, solving not for a deterministic strain field, but instead calculating the probability distribution of strain at each spatial coordinate, revealing a distribution over strain tensor functions.

\begin{figure}[H]
    \centering
    \includegraphics[scale=0.69]{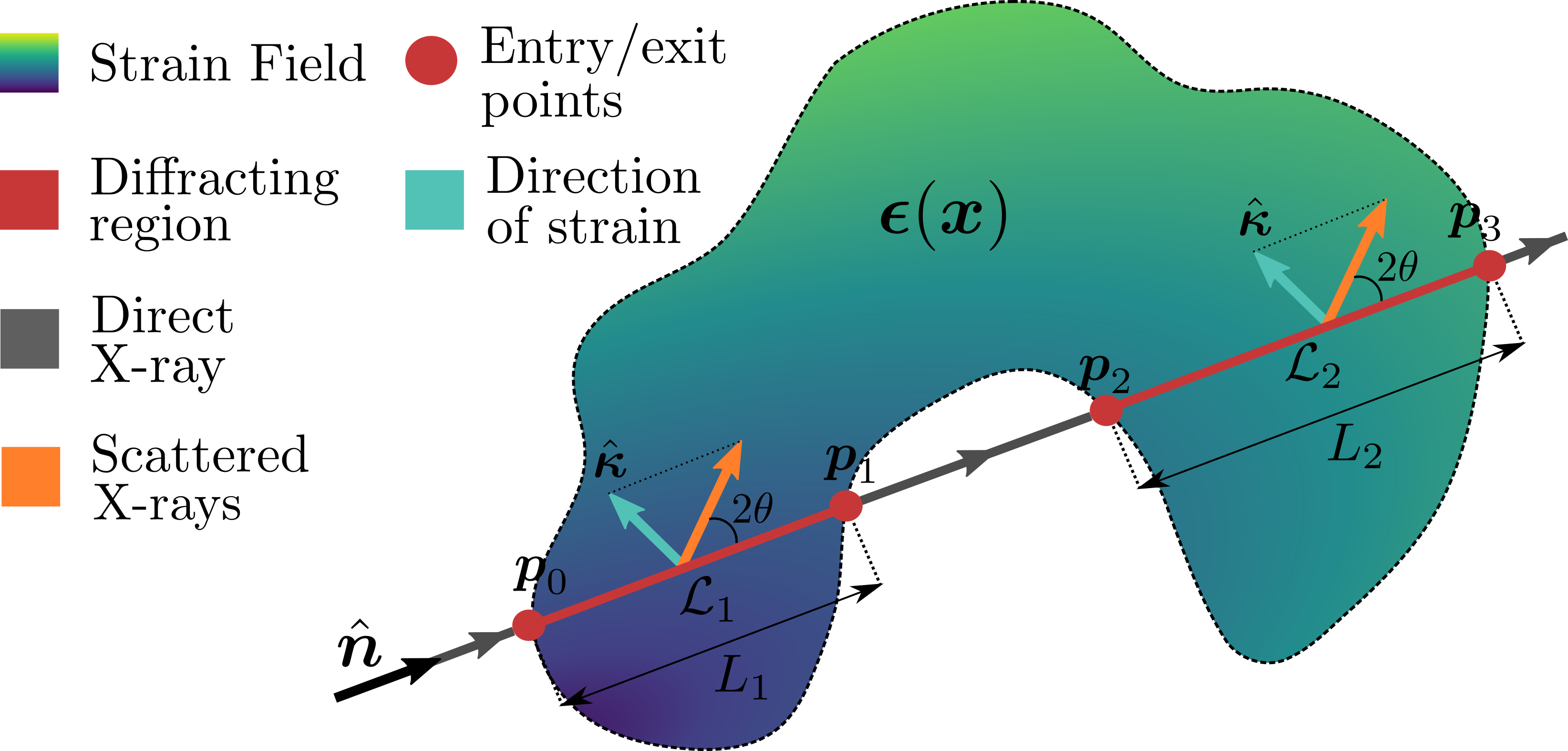}
    \caption{A single crystal under elastic deformation illuminated by an X-ray. Scattering takes place along the illuminated region \(\mathcal{L}=\mathcal{L}_1+\mathcal{L}_2\).}
    \label{fig:grain_and_ray}
\end{figure}

Before proceeding any further, it is useful to introduce a vector notation along with some geometrical quantities related to the integration path \(\mathcal{L}\) (Figure \ref{fig:grain_and_ray}). Continuing to work in the the \(\omega\)-coordinate system we define the column vectors
\begin{equation}
    \bar{\boldsymbol{\epsilon}} = \begin{bmatrix}\epsilon_{xx}\\\epsilon_{yy}\\\epsilon_{zz}\\\epsilon_{xy}\\\epsilon_{xz}\\\epsilon_{yz} \end{bmatrix},\quad \bar{\boldsymbol{\kappa}}=\begin{bmatrix}\kappa_{x}^2\\\kappa_{y}^2\\\kappa_{z}^2\\2\kappa_{x}\kappa_{y}\\2\kappa_{x}\kappa_{z}\\2\kappa_{y}\kappa_{z} \end{bmatrix},
\end{equation}
such that \(\bar{\boldsymbol{\kappa}}^T \bar{\boldsymbol{\epsilon}} = \hat{\boldsymbol{\kappa}}^T\boldsymbol{\epsilon}_{\omega}\hat{\boldsymbol{\kappa}}\) where \(\bar{\cdot}\) indicates flattening of a quantity to a column vector. Next, denoting the intersection points between X-ray beam and grain boundary by \(\boldsymbol{p}_0,\boldsymbol{p}_1,..,\boldsymbol{p}_M\) and letting the euclidean length of these illuminated regions be labelled \(L_i=||\boldsymbol{p}_i - \boldsymbol{p}_{i+1} ||_2\), we find, for measurement number \(j\), that
\begin{equation}
    y_{j} = e_j + \sum^{i=M-1}_{i=0} \dfrac{1}{L_i}\int^{L_i}_0 \bar{\boldsymbol{\kappa}}^T\bar{\boldsymbol{\epsilon}}(\boldsymbol{p}_i + \hat{\boldsymbol{n}}s) ds = e_j + \mathcal{M}_{j}\bar{\boldsymbol{\epsilon}},
\label{eq:vectorized_measurement_model}    
\end{equation}
where the symbol \(\mathcal{M}_j\) is shorthand for the integral operator corresponding to measurement number \(j\), \(s\) is a scalar and \(\hat{\boldsymbol{n}}\) is a unit vector along the X-ray beam. Considering a full measurement set, \(\boldsymbol{y}\), defined in \eqref{eq:y_full_measurement_set}, we introduce a compact notation,
\begin{equation}
    \boldsymbol{y} = \boldsymbol{\mathcal{M}}\bar{\boldsymbol{\epsilon}} + \boldsymbol{e}.
\end{equation}
where the \(\boldsymbol{\mathcal{M}}\) and \(\boldsymbol{e}\) are column vectors formed in analogy with \eqref{eq:y_full_measurement_set}.
\subsection{Gaussian Process regression}
A Gaussian Process is any stochastic process in which all subsets of a generated stochastic sequence of measurements form multivariate normal distributions \cite{rasmussen2003gaussian}. The regression procedure associated with a Gaussian Process, known as Gaussian Process regression, can be described in terms of basic statistical theorems and quantities. The central idea is to exploit that linear operators acting on normally distributed variables form again normal distributions. The goal is to arrive at the distribution of the Gaussian Process, which, for some spatial function \(f(\boldsymbol{x})\), describes the probability to find a value \(f\) at coordinate \(\boldsymbol{x}\) together with the covariance of \(f(\boldsymbol{x})\) with other spatial locations \(f(\boldsymbol{x}')\).

In the scanning-3DXRD case, we consider the measurement series, \(\boldsymbol{y}\),
generated by some underlying strain tensor field, \(\bar{\boldsymbol{\epsilon}}(\boldsymbol{x})\), and seek to calculate at each spatial coordinate, \(\boldsymbol{x}\), the probability distribution \(p(\boldsymbol{\bar{\epsilon}}(\boldsymbol{x})| \boldsymbol{y} )\), i.e, the probability to find a specified strain tensor  \(\bar{\boldsymbol{\epsilon}}\) at \(\boldsymbol{x}\) given the measurements \( \boldsymbol{y}\). As we will show, if we assume a Gaussian Process prior and Gaussian noise, this probability distribution is multivariate normal, and the covariance of strain at any two points, \(\mathbb{C}[\bar{\boldsymbol{\epsilon}}=\bar{\boldsymbol{\epsilon}}(\boldsymbol{x}),\bar{\boldsymbol{\epsilon}}'=\bar{\boldsymbol{\epsilon}}(\boldsymbol{x}')]\), together with the strain expected value, \(\mathbb{E}[\bar{\boldsymbol{\epsilon}}(\boldsymbol{x})]\) will be revealed by the regression.

If it is assumed that \(\bar{\boldsymbol{\epsilon}}(\boldsymbol{x})\) is normally distributed, 
\begin{equation}
    \bar{\boldsymbol{\epsilon}}(\boldsymbol{x}) \sim \mathcal{N}\bigg(\mathbb{E}[\bar{\boldsymbol{\epsilon}}],\mathbb{C}[\bar{\boldsymbol{\epsilon}},\bar{\boldsymbol{\epsilon}}']\bigg),
\end{equation}
it follows directly that \(\boldsymbol{y}\) is multivariate normal,
\begin{equation}
    \boldsymbol{y} \sim \mathcal{N}\bigg( \mathbb{E}[\boldsymbol{y}], \mathbb{C}[\boldsymbol{y},\boldsymbol{y}]\bigg),
    \label{eq:joint_prob_eps_I_compact}
\end{equation}
since it is a linear combination of the independent normal distributions \(\bar{\boldsymbol{\epsilon}}(\boldsymbol{x})\) and \(\boldsymbol{e}\). Considering then the joint distribution of \(\bar{\boldsymbol{\epsilon}}(\boldsymbol{x})\) and  \(\boldsymbol{y}\) we may calculate
\begin{equation}
     \begin{bmatrix}\bar{\boldsymbol{\epsilon}}\\ \boldsymbol{y}\end{bmatrix} \sim \mathcal{N}\Bigg(\begin{bmatrix}\boldsymbol{I}\\ \boldsymbol{\mathcal{M}}\end{bmatrix}\mathbb{E}[\bar{\boldsymbol{\epsilon}}], \begin{bmatrix}\mathbb{C}[\bar{\boldsymbol{\epsilon}},\bar{\boldsymbol{\epsilon}}']&\mathbb{C}[\bar{\boldsymbol{\epsilon}},\bar{\boldsymbol{\epsilon}}']\boldsymbol{\mathcal{M}}^T\\
     \boldsymbol{\mathcal{M}}\mathbb{C}[\bar{\boldsymbol{\epsilon}},\bar{\boldsymbol{\epsilon}}']& \boldsymbol{\mathcal{M}}\mathbb{C}[\bar{\boldsymbol{\epsilon}},\bar{\boldsymbol{\epsilon}}']\boldsymbol{\mathcal{M}}^T+\mathbb{C}[\boldsymbol{e},\boldsymbol{e}]\end{bmatrix}\Bigg),
     \label{eq:joint_prob_eps_I_expanded}
\end{equation}
where \(\boldsymbol{I}\) is an identity operator and it was used that \(\boldsymbol{y}\) is a linear transformation of two normally distributed variables \(\bar{\boldsymbol{\epsilon}}(\boldsymbol{x})\) and \(\boldsymbol{e}\). The joint probability of \eqref{eq:joint_prob_eps_I_expanded} now gives us the sought distribution, \(p( \bar{\boldsymbol{\epsilon}}(\boldsymbol{x})| \boldsymbol{y} )\), which is again normal. The calculation of its variance and expected value can be found by writing out \eqref{eq:joint_prob_eps_I_expanded} in analytical exponent form, with fixed \(\boldsymbol{y}\), and completing the exponent square. The closed form solution can be obtained
\begin{equation}
\begin{split}
    & \mathbb{E}[\bar{\boldsymbol{\epsilon}}|\boldsymbol{y}]= \mathbb{E}[\bar{\boldsymbol{\epsilon}}] + \mathbb{C}[\bar{\boldsymbol{\epsilon}},\bar{\boldsymbol{\epsilon}}']\boldsymbol{\mathcal{M}}^T\bigg(\boldsymbol{\mathcal{M}}\mathbb{C}[\bar{\boldsymbol{\epsilon}},\bar{\boldsymbol{\epsilon}}']\boldsymbol{\mathcal{M}}^T+\mathbb{C}[\boldsymbol{e},\boldsymbol{e}]\bigg)^{-1}\bigg(\boldsymbol{y} - \mathbb{E}[\boldsymbol{y}]\bigg),\\ &\mathbb{C}[\bar{\boldsymbol{\epsilon}},\bar{\boldsymbol{\epsilon}'}|\boldsymbol{y}] = \mathbb{C}[\bar{\boldsymbol{\epsilon}},\bar{\boldsymbol{\epsilon}}'] - \mathbb{C}[\bar{\boldsymbol{\epsilon}},\bar{\boldsymbol{\epsilon}}']\boldsymbol{\mathcal{M}}^T\bigg(\boldsymbol{\mathcal{M}}\mathbb{C}[\bar{\boldsymbol{\epsilon}},\bar{\boldsymbol{\epsilon}}']\boldsymbol{\mathcal{M}}^T+\mathbb{C}[\boldsymbol{e},\boldsymbol{e}]\bigg)^{-1}\boldsymbol{\mathcal{M}}\mathbb{C}[\bar{\boldsymbol{\epsilon}},\bar{\boldsymbol{\epsilon}}'].
\end{split}
\label{eq:general_posterior}
\end{equation}
Before any approximate or analytical solutions to the involved transformations of \(\mathbb{C}[\bar{\boldsymbol{\epsilon}},\bar{\boldsymbol{\epsilon}}']\) by \(\boldsymbol{\mathcal{M}}\) can be given, it remains first to specify the prior distribution of \(\bar{\boldsymbol{\epsilon}}(\boldsymbol{x})\).

\subsection{Equilibrium Prior}

Since the closed form solution of \eqref{eq:general_posterior} requires only that \(\bar{\boldsymbol{\epsilon}}(\boldsymbol{x})\) is normal, we are free to incorporate prior knowledge on \(\bar{\boldsymbol{\epsilon}}(\boldsymbol{x})\) by making a parametrisation of \(\bar{\boldsymbol{\epsilon}}(\boldsymbol{x})\) as linear transformations of some other underlying normal distributions. Since \(\bar{\boldsymbol{\epsilon}}(\boldsymbol{x})\) represents a linear elastic strain field and the scanning-3DXRD experiment is assumed to take place on a sample at rest, we expect that the accompanying stress field \(\bar{\boldsymbol{\sigma}}\) will be in static equilibrium. 
This can be expressed as a linear map
\begin{equation}
    \bar{\boldsymbol{\epsilon}}(\boldsymbol{x}) = \boldsymbol{H}\bar{\boldsymbol{\sigma}}(\boldsymbol{x}),
\end{equation}
where \(\boldsymbol{H}\) is an anisotropic compliance matrix that is orientation dependent, \(\boldsymbol{H}=\boldsymbol{H}(\boldsymbol{U})\approx \boldsymbol{H}(\boldsymbol{U}_0) \). The set of analytical functions \(\bar{\boldsymbol{\sigma}}(\boldsymbol{x})\) that satisfies balance of angular and linear momentum are known as the Beltrami Stress functions. These may be described as a linear map
\begin{equation}
    \bar{\boldsymbol{\sigma}}(\boldsymbol{x}) =\boldsymbol{\mathcal{B}}\bar{\boldsymbol{\Phi}}(\boldsymbol{x}),
\end{equation}
where \(\bar{\boldsymbol{\Phi}}(\boldsymbol{x})\) is a column vector holding six Beltrami stress functions which are required to be twice differentiable, and 
\begin{equation}
    \boldsymbol{\mathcal{B}} = \begin{bmatrix}
    0 & \frac{\partial^2}{\partial z^2} & \frac{\partial^2}{\partial y^2} & 0 & 0 & -2\frac{\partial^2}{\partial y \partial z} \\
    \frac{\partial^2}{\partial z^2} & 0 & \frac{\partial^2}{\partial x^2} & 0 & -2\frac{\partial^2}{\partial x \partial y} & 0 \\
    \frac{\partial^2}{\partial y^2} & \frac{\partial^2}{\partial x^2} & 0 & -2\frac{\partial^2}{\partial x \partial y} & 0 & 0 \\
    0 & 0 & -\frac{\partial^2}{\partial x \partial y} & -\frac{\partial^2}{\partial z^2} & \frac{\partial^2}{\partial y \partial z} & \frac{\partial^2}{\partial x \partial z} \\
    -\frac{\partial^2}{\partial y \partial z} & 0 & 0 & \frac{\partial^2}{\partial x \partial z} & \frac{\partial^2}{\partial x \partial y} & -\frac{\partial^2}{\partial x^2} \\
    0 & -\frac{\partial^2}{\partial x \partial z} & 0 & \frac{\partial^2}{\partial y \partial z} & -\frac{\partial^2}{\partial y^2} & \frac{\partial^2}{\partial x \partial y} \\
    \end{bmatrix}.
\end{equation}
We have, in total,
\begin{equation}
    \bar{\boldsymbol{\epsilon}}(\boldsymbol{x}) = \boldsymbol{H}\boldsymbol{\mathcal{B}}\bar{\boldsymbol{\Phi}}(\boldsymbol{x}),
\end{equation}
and must now make an assumption on the distribution of \(\bar{\boldsymbol{\Phi}}(\boldsymbol{x})\). Without any further prior knowledge we select a zero-mean normal distribution as
\begin{equation}
\begin{split}
    \mathbb{E}[\bar{\boldsymbol{\Phi}}]=\begin{bmatrix}0\\0\\0\\0\\0\\0\end{bmatrix},\quad
    \mathbb{C}[\bar{\boldsymbol{\Phi}},\bar{\boldsymbol{\Phi}}'] 
    = \begin{bmatrix}k_{1}&0&0&0&0&0\\
    0&k_{2}&0&0&0&0\\
    0&0&k_{3}&0&0&0\\
    0&0&0&k_{4}&0&0\\
    0&0&0&0&k_{5}&0\\
    0&0&0&0&0&k_{6}\\\end{bmatrix},
\end{split}
\end{equation}
where the covariance functions \(k_i  = k_i(\boldsymbol{x},\boldsymbol{x}')\) describe the spatial correlation of the field. In this work, we have used the stationary squared-exponential covariance function,
\begin{equation}
    k_i(\boldsymbol{x},\boldsymbol{x}') = \sigma_{i}^2\exp\bigg( \dfrac{-\boldsymbol{r}^T\boldsymbol{r}}{2\boldsymbol{l}_i^T\boldsymbol{l}_i} \bigg), \quad \boldsymbol{r} = \boldsymbol{x} - \boldsymbol{x}', \quad \boldsymbol{l}_i=\begin{bmatrix} l_{ix} & l_{iy} & l_{iz}
    \end{bmatrix}^T,
    \label{eq:squared_exp}
\end{equation}
introducing a smoothness assumption into the strain field reconstruction. The unknown hyperparameters defined by \(\boldsymbol{l}_i\) and \(\sigma_{i}\) are thus in total 6x4=24 in our case. These variables will be estimated through an initial optimisation process known as hyperparameter optimisation, we will return to how this is done later. First we show that the zero-mean \(prior\) assumption on the Beltrami stress functions, \(\bar{\boldsymbol{\Phi}}(\boldsymbol{x})\), does not imply that the \(posterior\) distribution of strain, \(\bar{\boldsymbol{\epsilon}}(\boldsymbol{x})\), will be zero-mean. This is realised upon examination of equation \eqref{eq:general_posterior} (a). Other selections for the prior mean are possible, however, when such additional prior information is unknown, a zero-mean selection is preferable for simplicity.

In total, these selections impose that (I) the strain field is in a point-wise static equilibrium and (II) that the strain field has a local spatial correlation to neighbouring points. The resulting prior distribution of strain is
\begin{equation}
    \bar{\boldsymbol{\epsilon}} \sim \mathcal{N}\bigg(\boldsymbol{0}, \boldsymbol{H}\boldsymbol{\mathcal{B}} \mathbb{C}[\bar{\boldsymbol{\Phi}},\bar{\boldsymbol{\Phi}}'] \boldsymbol{\mathcal{B}}^T\boldsymbol{H}^T\bigg).
\end{equation}

\subsection{Equilibrium Posterior Distribution}
With the prior information of equilibrium and spatial correlation now encoded into the strain field we may insert  
\begin{equation}
    \mathbb{C}[\bar{\boldsymbol{\epsilon}},\bar{\boldsymbol{\epsilon}}']=\boldsymbol{H}\boldsymbol{\mathcal{B}} \mathbb{C}[\bar{\boldsymbol{\Phi}},\bar{\boldsymbol{\Phi}}'] \boldsymbol{\mathcal{B}}^T\boldsymbol{H}^T
    \label{eq:post_strain_cov}
\end{equation}
into equation \eqref{eq:general_posterior} to arrive at a final expression in which only the hyperparameters remain to be estimated.
The covariance between measurements takes on the form
\begin{equation}
    \boldsymbol{\mathcal{M}}\boldsymbol{H}\boldsymbol{\mathcal{B}} \mathbb{C}[\bar{\boldsymbol{\Phi}},\bar{\boldsymbol{\Phi}}'] \boldsymbol{\mathcal{B}}^T\boldsymbol{H}^T\boldsymbol{\mathcal{M}}^T,
\end{equation}
which involves, through the mappings \(\boldsymbol{\mathcal{M}}\), a double integral over the two times partially differentiated squared-exponential in \eqref{eq:squared_exp}. 
The solution to this double line integral is intractable, although some work has been done to show that for \(l_x=l_y=l_z\) it can be analytically reduced to a single integral \cite{hendriks2019implementation}. However, the numerical integration remains too computationally costly for practical use. 
This motivates the use of an approximation scheme on a reduced basis for which closed form solutions to all involved quantities of \eqref{eq:general_posterior} are again recovered \cite{Jidling2018}.

\subsection{Finite Basis Approximations}
Decomposing \eqref{eq:squared_exp} onto a Fourier basis,
\begin{equation}
    \varphi_{ik}(\boldsymbol{x}) = \dfrac{1}{L_xL_yL_z}\sin(\lambda_{xik}(x+L_x))\sin(\lambda_{yik}(y+L_y))\sin(\lambda_{zik}(z+L_z))
    \label{eq:fourier_basis}
\end{equation}
we find that
\begin{equation}
    k_i(\boldsymbol{x},\boldsymbol{x}') \approx \sum^{k=m}_{k=1} \varphi_{ik}(\boldsymbol{x})s_{ik}\varphi_{ik}(\boldsymbol{x}')=  \boldsymbol{\varphi}^T_{i} \boldsymbol{s}_{i} \boldsymbol{\varphi}'_{i},
    \label{eq:decomposed_kernel}
\end{equation}
where \(\boldsymbol{s}_i\) is a diagonal matrix of basis coefficients, \(s_{ik}\), which are the spectral densities of \eqref{eq:squared_exp}. Specifically it is possible to show \cite{solin2020hilbert} that the \(k\):th spectral density is,
\begin{equation}
    s_{ik} = \sigma^2_i(2\pi)^{\frac{2}{3}}l_{ix}l_{iy}l_{iz}\exp\bigg( -\frac{1}{2}(l^2_{ix}\lambda^2_{xik} + l^2_{iy}\lambda^2_{yik} + l^2_{iz}\lambda^2_{zik}) \bigg).
    \label{eq:spectral_density}
\end{equation}
With the vector notation
\begin{equation}
    \boldsymbol{\phi} = 
    \begin{bmatrix}\boldsymbol{\varphi}_1\\
    \boldsymbol{\varphi}_2\\
    \boldsymbol{\varphi}_3\\
    \boldsymbol{\varphi}_4\\
    \boldsymbol{\varphi}_5\\
    \boldsymbol{\varphi}_6\end{bmatrix}, \quad     \boldsymbol{S} = 
    \begin{bmatrix}\boldsymbol{s}_1&\boldsymbol{0}&\boldsymbol{0}&\boldsymbol{0}&\boldsymbol{0}&\boldsymbol{0}\\
    \boldsymbol{0}&\boldsymbol{s}_2&\boldsymbol{0}&\boldsymbol{0}&\boldsymbol{0}&\boldsymbol{0}\\
    \boldsymbol{0}&\boldsymbol{0}&\boldsymbol{s}_3&\boldsymbol{0}&\boldsymbol{0}&\boldsymbol{0}\\
    \boldsymbol{0}&\boldsymbol{0}&\boldsymbol{0}&\boldsymbol{s}_4&\boldsymbol{0}&\boldsymbol{0}\\
    \boldsymbol{0}&\boldsymbol{0}&\boldsymbol{0}&\boldsymbol{0}&\boldsymbol{s}_5&\boldsymbol{0}\\
    \boldsymbol{0}&\boldsymbol{0}&\boldsymbol{0}&\boldsymbol{0}&\boldsymbol{0}&\boldsymbol{{s}_6}
    \end{bmatrix}, 
\end{equation}
where \(\boldsymbol{0}\) is a matrix of zeros, we find the approximate covariance,
\begin{equation}
    \mathbb{C}[\bar{\boldsymbol{\Phi}},\bar{\boldsymbol{\Phi}}'] = \boldsymbol{\phi}^T \boldsymbol{S} \boldsymbol{\phi}'.
    \label{eq:spectral_decomp_covariance}
\end{equation}
Insertion of \eqref{eq:spectral_decomp_covariance} into \eqref{eq:post_strain_cov} now yields
\begin{equation}
    \mathbb{C}[\bar{\boldsymbol{\epsilon}},\bar{\boldsymbol{\epsilon}}']=\boldsymbol{H}\boldsymbol{\mathcal{B}} \boldsymbol{\phi}^T \boldsymbol{S} \boldsymbol{\phi}' \boldsymbol{\mathcal{B}}^T\boldsymbol{H}^T.
    \label{eq:post_approx_strain_cov}
\end{equation}
Introducing the quantities
\begin{equation}
    \boldsymbol{\phi}_{\epsilon} = \boldsymbol{H}\boldsymbol{\mathcal{B}}\boldsymbol{\phi}^T, \quad \boldsymbol{\phi}_{y} = \boldsymbol{\mathcal{M}}\boldsymbol{\phi}_{\epsilon},
\end{equation}
we may finally arrive at the approximate posterior mean and covariance of strain using \eqref{eq:general_posterior}, as
\begin{equation}
\begin{split}
    & \mathbb{E}[\bar{\boldsymbol{\epsilon}}|\boldsymbol{y}]= \mathbb{E}[\bar{\boldsymbol{\epsilon}}] + \boldsymbol{\phi}_{\epsilon}\boldsymbol{S}\boldsymbol{\phi}_{y}^T\bigg( \boldsymbol{\phi}_{y}\boldsymbol{S}\boldsymbol{\phi}_{y}^T +\mathbb{C}[\boldsymbol{e},\boldsymbol{e}]\bigg)^{-1}\bigg(\boldsymbol{y} - \mathbb{E}[\boldsymbol{y}]\bigg),\\ &\mathbb{C}[\bar{\boldsymbol{\epsilon}},\bar{\boldsymbol{\epsilon}}|\boldsymbol{y}] = \boldsymbol{\phi}_{\epsilon}\boldsymbol{S}\boldsymbol{\phi}_{\epsilon}^T - \boldsymbol{\phi}_{\epsilon}\boldsymbol{S}\boldsymbol{\phi}_{y}^T\bigg( \boldsymbol{\phi}_{y}\boldsymbol{S}\boldsymbol{\phi}_{y}^T +\mathbb{C}[\boldsymbol{e},\boldsymbol{e}]\bigg)^{-1} \boldsymbol{\phi}_y\boldsymbol{S}\boldsymbol{\phi}^T_{\epsilon}.
\end{split}
\label{eq:approximate_posterior}
\end{equation}
The computational complexity can be further reduced by algebraically rearranging this equation to avoid forming the covariance matrices \cite{rasmussen2003gaussian}, resulting in
\begin{equation}
\begin{split}
    &\mathbb{E}[\bar{\boldsymbol{\epsilon}}|\boldsymbol{y}] = \mathbb{E}[\bar{\boldsymbol{\epsilon}}] + \boldsymbol{\phi}_{\epsilon}\bigg(\boldsymbol{\phi}_{y}^T\mathbb{C}[\boldsymbol{e},\boldsymbol{e}]^{-1}\boldsymbol{\phi}_{y} + \boldsymbol{S}^{-1}\bigg)^{-1}\boldsymbol{\phi}_{y}^T\mathbb{C}[\boldsymbol{e},\boldsymbol{e}]^{-1}\bigg(\boldsymbol{y} - \mathbb{E}[\boldsymbol{y}]\bigg), \\
    &\mathbb{C}[\bar{\boldsymbol{\epsilon}},\bar{\boldsymbol{\epsilon}}|\boldsymbol{y}] = \boldsymbol{\phi}_{\epsilon}\bigg(\boldsymbol{\phi}_{y}^T\mathbb{C}[\boldsymbol{e},\boldsymbol{e}]^{-1}\boldsymbol{\phi}_{y} + \boldsymbol{S}^{-1}\bigg)^{-1}\boldsymbol{\phi}_{\epsilon}^T.
\end{split}
\label{eq:approx_sol}
\end{equation}
Here, the inverses \(\boldsymbol{S}^{-1}\) and \(\mathbb{C}[\boldsymbol{e},\boldsymbol{e}]^{-1}\) can be trivially computed, as the matrices are diagonal.
For \(m < N\), this reduces the computational complexity to \(\mathcal{O}(Nm^2)\) from \(\mathcal{O}(N^3)\) required for the inverse in \eqref{eq:general_posterior} and \eqref{eq:approximate_posterior}.
A numerically stable and efficient algorithm for solving these equations using the QR decomposition can be found within \citep{hendriks2019robust} together with analytical expressions for the various integral mappings \(\boldsymbol{\mathcal{M}}\). 

As \(m \to \infty\) the approximate solution \eqref{eq:approx_sol} approaches the exact solution \eqref{eq:general_posterior} \cite{solin2020hilbert}. In practice, however, we must select a finite \(m\), leading to \eqref{eq:squared_exp} being used in approximate form. To make a selection of frequencies, \(\lambda_{xik},\lambda_{yik},\lambda_{zik}\), in \eqref{eq:decomposed_kernel} use can be made of \eqref{eq:spectral_density}. In this work, we have selected the basis frequencies on an equidistant grid in (\(\lambda_{xik},\lambda_{yik},\lambda_{zik}\))-space such that spectral densities were above a minimum threshold i.e. we aim to achieve a desired coverage of the spectral density function. Specifically, we select
\begin{equation}
\begin{split}
    &\lambda_{xik} = \Delta\lambda_{xki} g_{xki}, \quad L_x = \dfrac{\pi}{2\Delta\lambda_{xki}},\quad \Delta\lambda_{xki} = 0.7 l_{ix},\\
    &\lambda_{yik} = \Delta\lambda_{yki} g_{yki}, \quad L_y = \dfrac{\pi}{2\Delta\lambda_{yki}},\quad \Delta\lambda_{yki} = 0.7 l_{iy},\\
    &\lambda_{zik} = \Delta\lambda_{zki} g_{zki}, \quad L_x = \dfrac{\pi}{2\Delta\lambda_{zki}},\quad \Delta\lambda_{zki} = 0.7 l_{iz},\\
    &g_{xki}^2 + g_{xki}^2 + g_{xki}^2 \leq \sqrt{5}
\end{split}
\end{equation}
where (\(g_{xki},g_{yki},g_{zki}\)) are positive integers such that (\(\Delta\lambda_{xki}g_{xki}, \Delta\lambda_{yki}g_{yki}, \Delta\lambda_{zki}g_{zki}\)) defines equidistant grid points excluding the origin. This selection results in a total of \(m\)=38 used basis functions.
To complete the regression scheme, we must now discuss the selection of hyperparameters, which at this stage, are the only unknowns in the formulation. 

\subsection{Hyperparameter selection}
Hyperparameters for the posterior conditional distribution can be determined through optimisation \cite{rasmussen2003gaussian}. Typically, this is done by maximising either the log marginal likelihood or using a cross-validation approach and maximising the \emph{out-of-sample} log likelihood, i.e. the likelihood of observing a set of measurements not used in the regression, $\tilde{\boldsymbol{y}}$. 
Following the work by \citep{gregg2020radial}, which demonstrates that maximising the out-of-sample log likelihood yields better results for line integral measurements, we determine the hyperparameters by solving
\begin{equation}
\begin{split}
    \Theta_* &=  \argmax_\Theta \log p_\Theta(\tilde{\boldsymbol{y}} | \boldsymbol{y})  \\
    &= \argmax_\Theta -0.5\log\det\mathbb{C}[\tilde{\boldsymbol{y}},\tilde{\boldsymbol{y}}|\boldsymbol{y}] - 0.5(\tilde{\boldsymbol{y}} - \mathbb{E}[\tilde{\boldsymbol{y}}|\boldsymbol{y}])^T\mathbb{C}[\tilde{\boldsymbol{y}},\tilde{\boldsymbol{y}}|\boldsymbol{y}]^{-1}(\tilde{\boldsymbol{y}} - \mathbb{E}[\tilde{\boldsymbol{y}}|\boldsymbol{y}]).
\end{split}
\end{equation}
By extension of \eqref{eq:approx_sol}, we have that
\begin{equation}
\begin{split}
    &\mathbb{E}[\bar{\tilde{\boldsymbol{y}}}|\boldsymbol{y}] = \mathbb{E}[\tilde{\boldsymbol{y}}] + \boldsymbol{\phi}_{\tilde{y}}\left(\boldsymbol{\phi}_{y}^T\mathbb{C}[\boldsymbol{e},\boldsymbol{e}]^{-1}\boldsymbol{\phi}_{y} + \boldsymbol{S}^{-1}\right)^{-1}\boldsymbol{\phi}_{y}^T\mathbb{C}[\boldsymbol{e},\boldsymbol{e}]^{-1}\left(\boldsymbol{y} - \mathbb{E}[\boldsymbol{y}]\right), \\
    &\mathbb{C}[\tilde{\boldsymbol{y}},\tilde{\boldsymbol{y}}|\boldsymbol{y}] = \boldsymbol{\phi}_{\tilde{y}}\left(\boldsymbol{\phi}_{y}^T\mathbb{C}[\boldsymbol{e},\boldsymbol{e}]^{-1}\boldsymbol{\phi}_{y} + \boldsymbol{S}^{-1}\right)^{-1}\boldsymbol{\phi}_{\tilde{y}}^T + \mathbb{C}[\boldsymbol{e},\boldsymbol{e}],
\end{split}
\end{equation}
and \(\Theta\) is a vector holding the hyper parameters introduced in \eqref{eq:squared_exp}.

It is noted that it is not essential that a global optima is found in this procedure, in fact, in many cases, setting the hyperparameters to some reasonable fixed values may produce excellent reconstructions. In the case of scanning-3DXRD we have found that setting the hyperparameters uniformly to the grain diameter gives reasonable results and can serve as a good inital guess for optimisation.

\section{Validation}\label{sec:Validation}
To validate the presented regression method we have generated simulated scanning-3DXRD data using a previously developed algorithm \cite{HenningssonThesis}. This tool has been used with success in the past (c.f \citep{Hektor2019} and \citep{Henningsson2020}) and can provide an understanding of the limitations and benefits of scanning-3DXRD reconstruction methods. Briefly, the simulation input is specified as a set of cubic single crystal voxels featuring individual strains and orientations together with an experimental setup. We refer the reader to \citep{HenningssonThesis} for additional details on the simulation algorithm with an undocumented implementation available via \url{https://github.com/FABLE-3DXRD/S3DXRD/}. Strain reconstructions from generated diffraction data were compared to ground truth input strain as well as an additional reconstruction method described in \citep{Henningsson2020}. This reconstruction method, previously referred to as \(algebraic\) \(strain\) \(refinement\) (ASR), uses a voxel basis for strain reconstruction and can, in short, be described as solving a global WLSQ problem. This least-squares approach operates from the same average directional strain data as the presented GP method.

\subsection{Single Crystal Simulation Test-case}\label{subsec:Single Crystal Simulation Test-case}
\begin{figure}[H]
\centering
\includegraphics[scale=0.5]{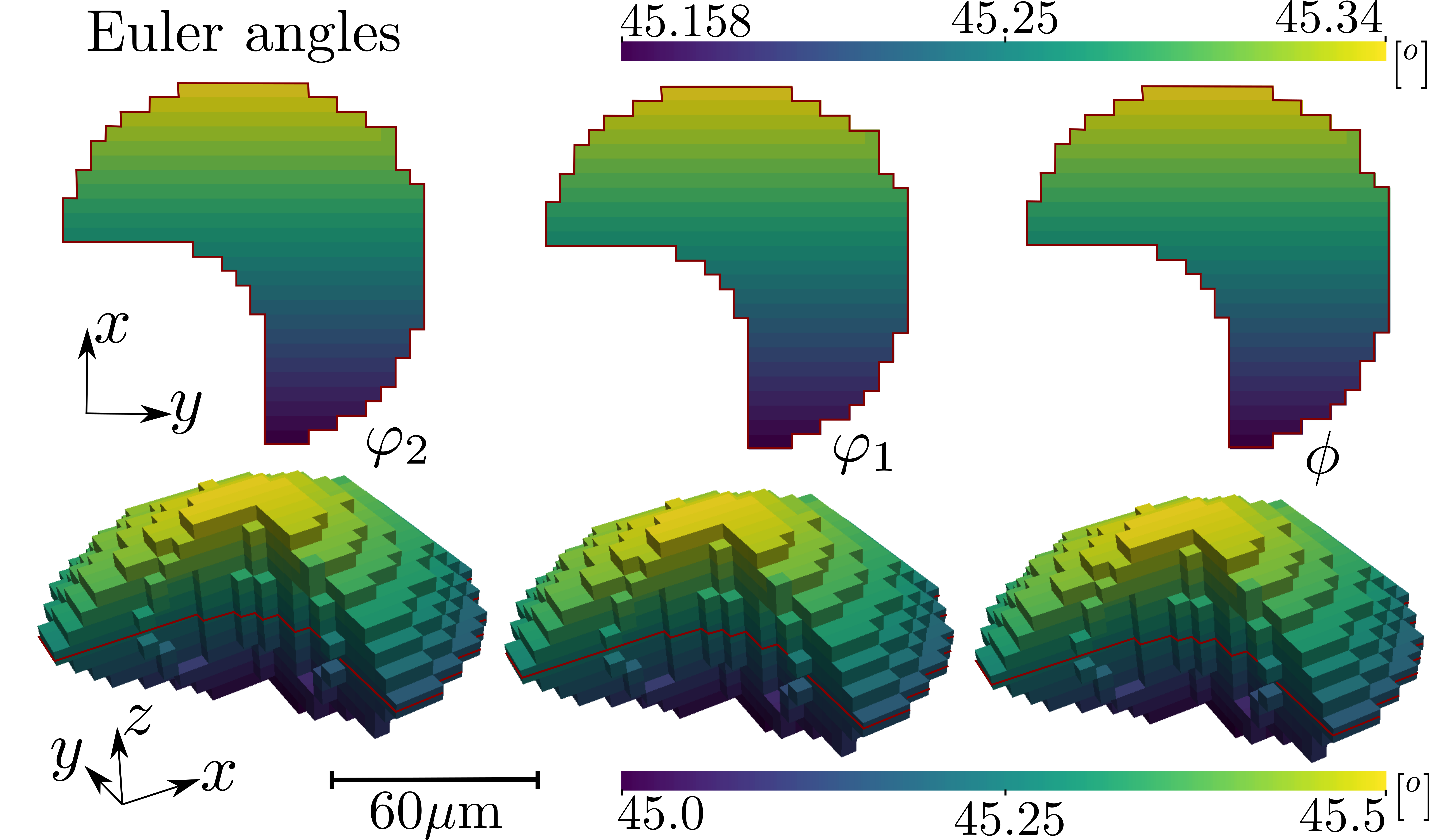}
\caption{Grain topology input for diffraction simulation coloured by corresponding input Euler angle field in units of degrees. The top row represents central cuts through the below 3D renderings as indicated by the the red lines.}
\label{fig:euler} 
\end{figure}

Diffraction from a tin (Sn) grain subject to a nonuniform strain tensor field has been simulated for the non-convex grain topology depicted in Figure \ref{fig:euler}.
The grain was assigned an orientation field by introducing linear gradients in the three Euler (Bunge notation) angles, \(\varphi_1,\Phi,\varphi_2\), as 
\begin{equation}
    \varphi_1 = \Phi = \varphi_2 = \dfrac{\pi}{180}\bigg(45 + \dfrac{x}{130v} + \dfrac{z}{24v} \bigg),
\end{equation}
where \(v=5\mu\)m was the used voxel size and the grain origin was set at the grain centroid in the \(x\)-\(y\)-plane and at the bottom edge of the grain in \(z\) (Figure \ref{fig:euler}). 

\begin{figure}[H]
\centering
\includegraphics[scale=0.5]{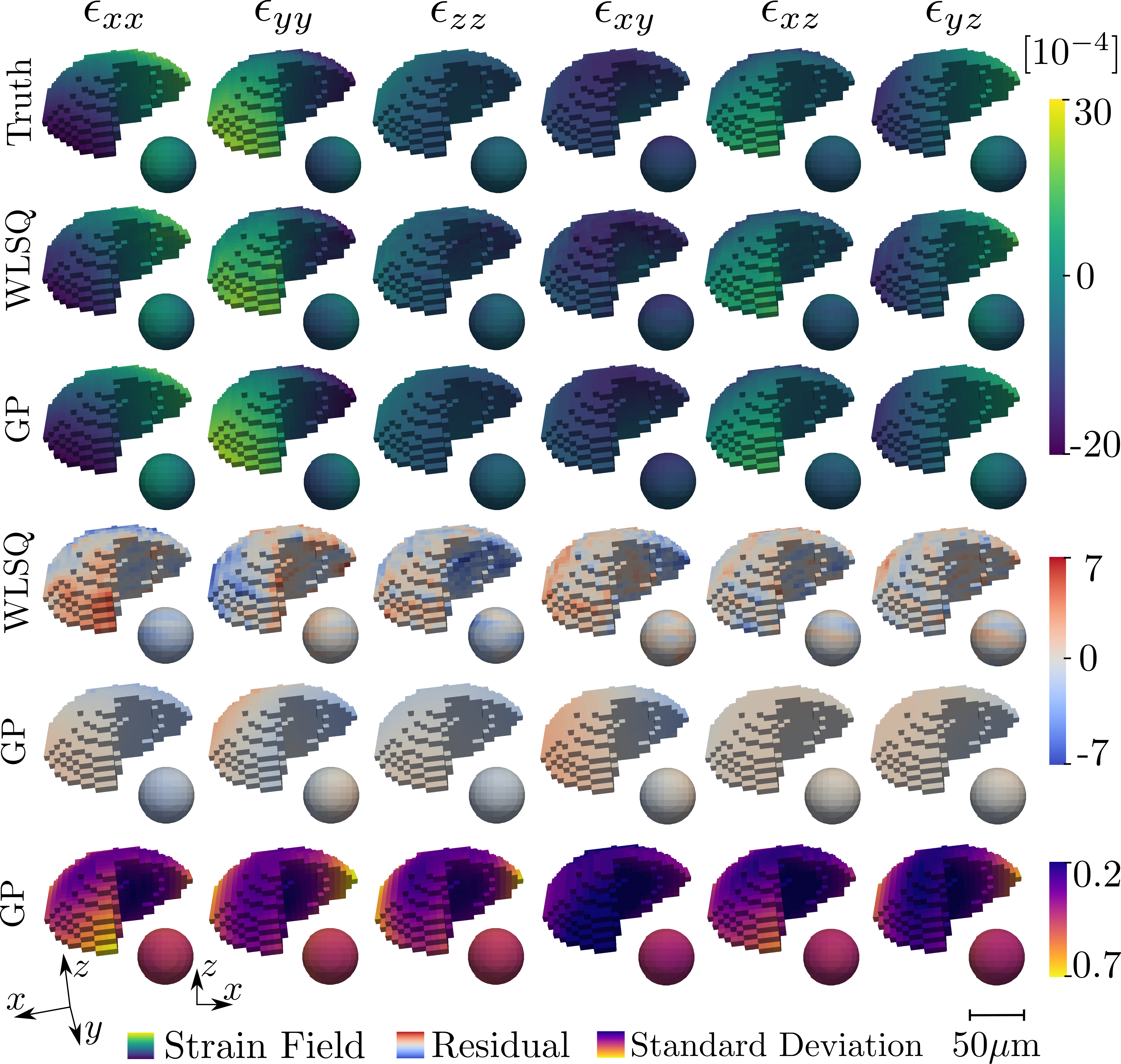}
\caption{3D rendering of strain reconstructions for a weighted least squares (WLSQ) and Gaussian Process (GP) regression approach. The top row defines the simulation ground truth as described in in equation \eqref{eq:maxwell_simulation_strain} with each column featuring a different strain component. The surface of the voxelated grain is presented together with a pulled out interior spherical cut centred at the grain centroid with diameter of 50\(\mu\)m. The corresponding coordinate systems are depicted in the bottom left of the figure. Three separate colormaps have been assigned to enhance contrast for the various fields, however, units of strain remain the same across plots [\(\times\)10\(^{-4}\)]. The residual field is defined as the difference between ground truth and reconstructed strain field. }
\label{fig:maxwell_simulation} 
\end{figure}

The strain field was defined by a set of Maxwell stress functions, which are a subset of the more general class of Beltrami stress functions,
\begin{equation}
    \bar{\boldsymbol{\Phi}} = 
    \begin{bmatrix}
    A(x,y,z)&B(x,y,z)&C(x,y,z)&0&0&0
    \end{bmatrix}^T.
\end{equation}
To achieve a relatively simple, but not trivial, strain field the functions \(A,B\) and \(C\) where selected as a cubic polynomial
\begin{equation}
\begin{split}
    &A=B=C=\\
    & \rho_1(x-t_x)^3 + \rho_2(y-t_y)^3 + \rho_3(z-t_z)^3 + \rho_4(x-t_x)(y-t_y)(z-t_z).
\end{split}
\end{equation}
The stress was converted to strain by the elastic compliance matrix, \(\boldsymbol{C}\), as
\begin{equation}
    \begin{bmatrix}
         \epsilon_{xx}  \\
         \epsilon_{yy}  \\
         \epsilon_{zz}  \\
         \epsilon_{xy}  \\
         \epsilon_{xz}  \\
         \epsilon_{yz}
    \end{bmatrix} = \boldsymbol{C}^{-1}\boldsymbol{\mathcal{B}}\bar{\boldsymbol{\Phi}}=
    \boldsymbol{C}^{-1}
    \begin{bmatrix}
        6\rho_3(z-t_z) + 6\rho_2(y-t_y)  \\
        6\rho_1(x-t_x) + 6\rho_3(z-t_z)  \\
        6\rho_2(y-t_y) + 6\rho_1(x-t_x)  \\
        \rho_4(t_z-z)  \\
        \rho_4(t_y-y)  \\
        \rho_4(t_x-x)
    \end{bmatrix}.
    \label{eq:maxwell_simulation_strain}
\end{equation}
Numerical values of the constants \(\rho_1,\rho_2,\rho_3,\rho_4,t_x,t_y,t_z\) are presented in Table \ref{tab:simulation_maxwell_parameters}
\begin{table}[H]
 \caption{Strain field parameters for diffraction simulation in units of \(\mu\)m.}
 \begin{tabular}{ccccccc}
  \(\rho_1\) & \(\rho_2\) & \(\rho_3\) & \(\rho_4\) & \(t_x\) & \(t_y\) & \(t_z\) \\
  \hline
  \(100\) &  \(100\) &  \(100\) & \(1000\) & \(10\) & \(10\) & \(0\)
\end{tabular}
\label{tab:simulation_maxwell_parameters}
\end{table}
The elasticity matrix for single crystal tin (Sn) was taken from \citep{Darbandi2013} (Table \ref{tab:Sn_elasticity_constants}) and converted from Voigt notation to the used strain vector notation.
\begin{table}[H]
 \caption{Elasticity constants for single crystal tin (Sn) in units of GPa converted from voigt notation as given in \citep{Darbandi2013}}
 \begin{tabular}{ccccccccc}
  \(C_{11}\) & \(C_{22}\) & \(C_{33}\) & \(C_{44}\) & \(C_{55}\) & \(C_{66}\) & \(C_{12}\) & \(C_{13}\)& \(C_{23}\)\\
  \hline
  72.3 & 72.3 & 88.4 & 48.0 & 44.0 & 44.0 & 59.4 & 35.8 & 35.8    
\end{tabular}
\label{tab:Sn_elasticity_constants}
\end{table}
Parameters presented in Table \ref{tab:simul_params} where used to define the experimental setup of the simulation. 
\begin{table}[H]
 \caption{Experimental parameters used in single grain simulation, corresponding to the results presented in Figure \ref{fig:maxwell_simulation}}
\begin{tabular}{ll}
Wavelength & 0.22\AA \\
Sample to detector distance  &   163mm  \\
Detector pixel size  &   50\(\mu\)m\(\times\)50\(\mu\)m    \\
Detector dimensions & 2048\(\times\)2048 pixels \\
Beam size  &   5\(\mu\)m\(\times\)5\(\mu\)m    \\
\(\omega\) rotation interval  &   [0\(^o\), 180\(^o\)]\\
\(\Delta \omega\) step length  &   1\(^o\)     \\
\end{tabular}
\label{tab:simul_params}
\end{table}
The unit cell in Table \ref{tab:lattice} was used to define a strain free lattice state.
\begin{table}[H]
\caption{Relaxed reference lattice.}
\begin{tabular}{c c c c c c}
\(a\) & \(b\) & \(c\) & \(\alpha\) & \(\beta\) & \(\gamma\) \\
5.81127 \AA & 5.81127 \AA & 3.17320 \AA &  90.0\(^o\) &  90.0\(^o\) &  90.0\(^o\)
\end{tabular}
\label{tab:lattice}
\end{table}
The generated diffraction patterns where analysed leading to a grain topology map together with a per slice average orientation. The diffraction data were then converted to average directional strains, as described in \ref{sec:Measurement Model}, and input to the WLSQ and GP reconstruction methods. The final reconstructed strain tensor fields are illustrated together with simulation ground truth and residual fields in Figure \ref{fig:maxwell_simulation}. 

\begin{figure}[H]
    \centering
    \includegraphics[scale=0.4]{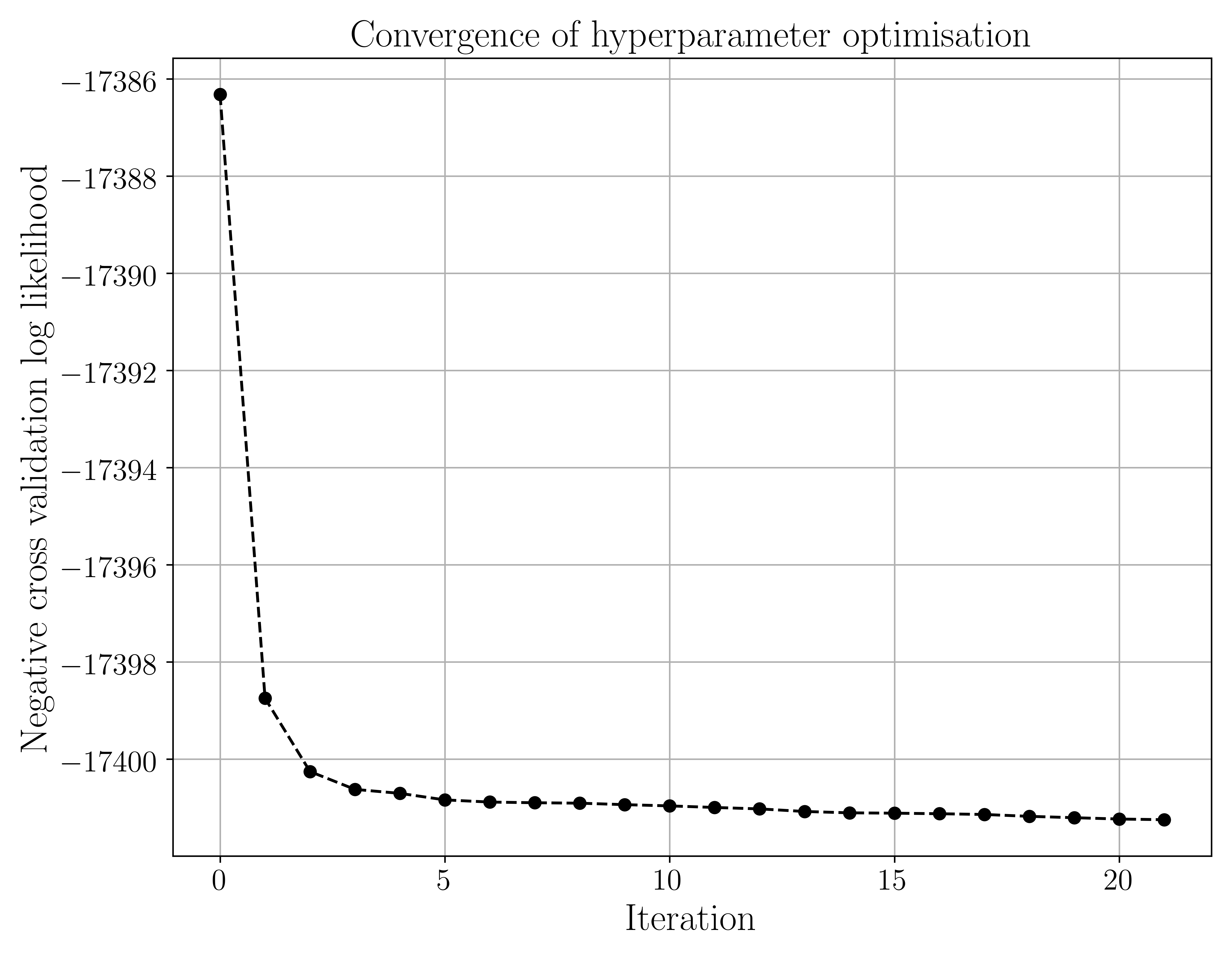}
    \caption{Negative cross validation log likelihood reduction during hyperparameter optimisation for the simulated grain presented in Figure \ref{fig:euler} and \ref{fig:maxwell_simulation}. Optimisation was conducted using the L-BFGS-B algorithm as implemented in scipy \cite{scipy} with a maximum of 10 line-search steps per iteration. Gradients where computed using automatic differentiation as implemented in torch.}
    \label{fig:hp_convergence_simulated_grain}
\end{figure}

Hyperparameters were optmised using the L-BFGS-B algorithm, as implemented in scipy \cite{scipy}, with a maximum of 10 line-search steps per iteration. Gradients where computed using automatic differentiation as implemented in PyTorch \cite{PyTorch}. In the first optimisation iteration all hyperparameters were uniformly set to the grain radius. Convergence of the optimisation is displayed in \ref{fig:hp_convergence_simulated_grain}. The smoothness constraints for the WLSQ in \(x\)-\(y\)-plane were set to 2.5\(\times\)10\(^{-4}\), limiting the maximum absolute difference in each strain tensor component between two neighbouring voxels (see \citep{Henningsson2020} for further details).

\begin{table}[H]
 \caption{Root mean squared errors, mean absolute errors and maximum absolute errors for the residual fields presented in Figure \ref{fig:maxwell_simulation}. The result of the Gaussian Process regression (GP) can be compared to the weighted least squares fit (WLSQ). Values are unit less (strain) and in the same scale (10\(^{-4}\)) as in Figure \ref{fig:maxwell_simulation}.}
 \begin{tabular}{ccccccc}
    \hline
    Strain &  \multicolumn{2}{c}{Root Mean Squared Error} & \multicolumn{2}{c}{Mean Absolute Error}& \multicolumn{2}{c}{Maximum Absolute Error}\\
    \hline
     &  GP &  WLSQ &  GP &  WLSQ &  GP &  WLSQ \\
    \(\epsilon_{xx}\) & 1.322 & 2.076 &  1.101 & 1.737 & 2.791       & 7.42\\
    \(\epsilon_{yy}\) & 1.042 & 1.371 &  0.846 & 0.999 & 3.856       & 6.094\\
    \(\epsilon_{zz}\) & 0.887 & 1.489 &  0.769 & 1.157 & 1.914       & 6.736\\
    \(\epsilon_{xy}\) & 1.122 & 1.511 &  0.955 & 1.172 & 2.778       & 6.306\\
    \(\epsilon_{xz}\) & 0.24  & 1.04  &  0.198 & 0.798 & 1.506       & 4.85\\ 
    \(\epsilon_{yz}\) & 0.48  & 0.958 &  0.399 & 0.742 & 1.34        & 4.652\\
    \hline
\end{tabular}
\label{tab:rmse_and_mae}
\end{table}

\begin{figure}[H]
    \centering
    \includegraphics[scale=0.5]{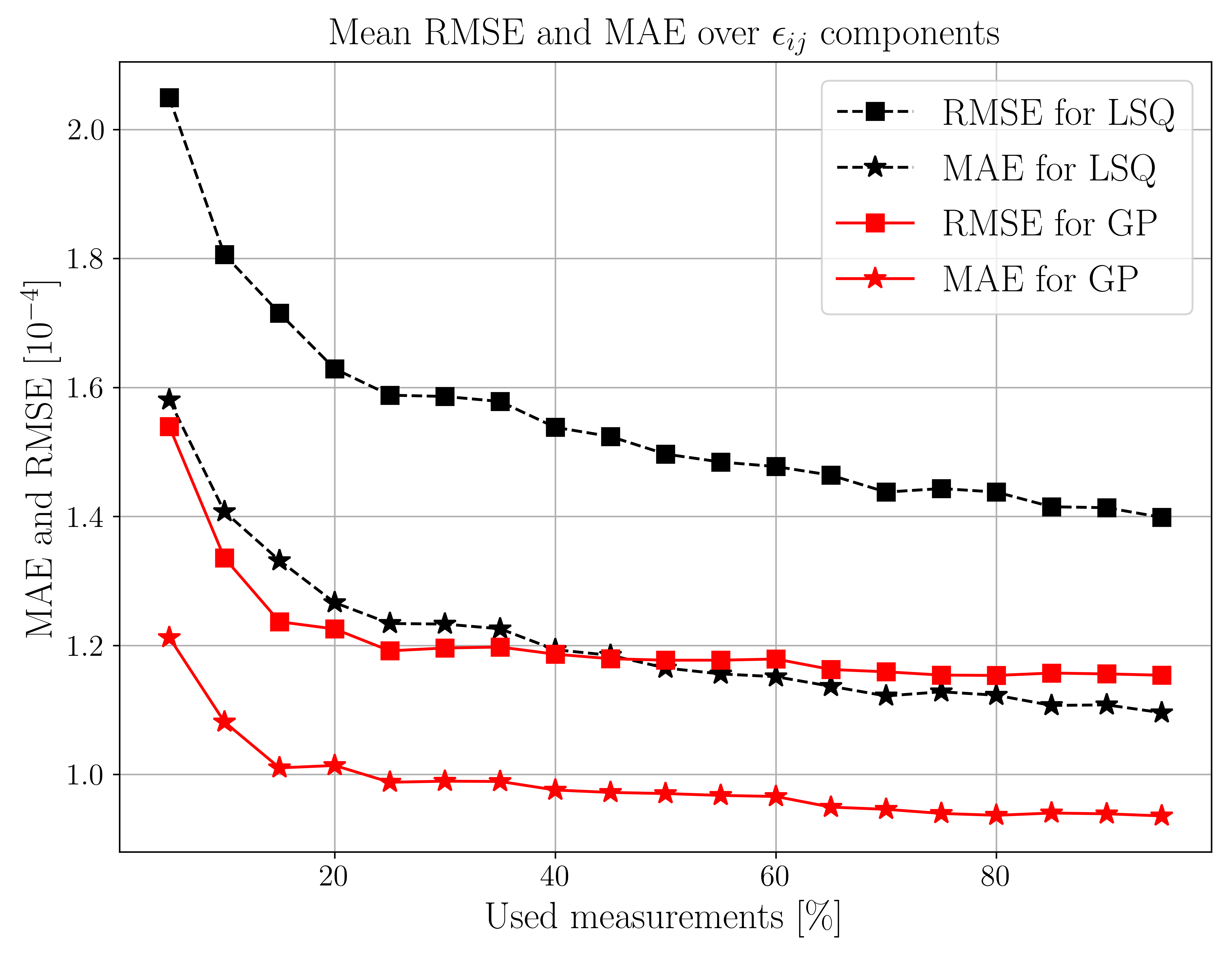}
    \caption{Average root mean squared error (squares) and absolute mean error (stars) for the simulated grain presented in Figure \ref{fig:euler} and \ref{fig:maxwell_simulation} as a function of used percentage of measurements. The performance of the Gaussian Process regression (Red filled lines) can be compared to that of the weighted least squares (black dashed lines). The RMSE and MAE were computed from the residual fields and averaged over the six reconstructed strain components to produce a scalar measure per reconstruction. Each point in the plot corresponds to a full 3D strain reconstruction using a random subset of the measured data.}
    \label{fig:convergence_study_rmse_and_mae}
\end{figure}

To asses how well the two methods (WSLQ and GP) utilises data, the average MAE and RMSE of reconstructed strain fields, as a function of the number of input measurement integrals, has been investigated. To avoid repeated hyperparameter optimisation and, at the same time, not benefit the GP reconstruction unfairly, the hyperparameters for the GP were set uniformly to the grain diameter. Although, this most probably represents a worse reconstruction than achievable by the GP method, it suffices to illustrate the elevated performance of the method when the number of available measurements is low. Measurements where permuted randomly and inputted to the WLSQ and GP reconstruction in steps of 5\% as indicated by the markers in Figure \ref{fig:convergence_study_rmse_and_mae}. The resulting MAE and RMSE for the reconstructed residual fields where computed and averages over the six strain components was formed. The performance as a function of input measurements can be assessed by visual inspection of Figure \ref{fig:convergence_study_rmse_and_mae}.

\subsection{Embedded tin (Sn) Grain}\label{subsec:Embedded tin (Sn) Grain}
\begin{figure}[H]
    \centering
    \includegraphics[scale=0.5]{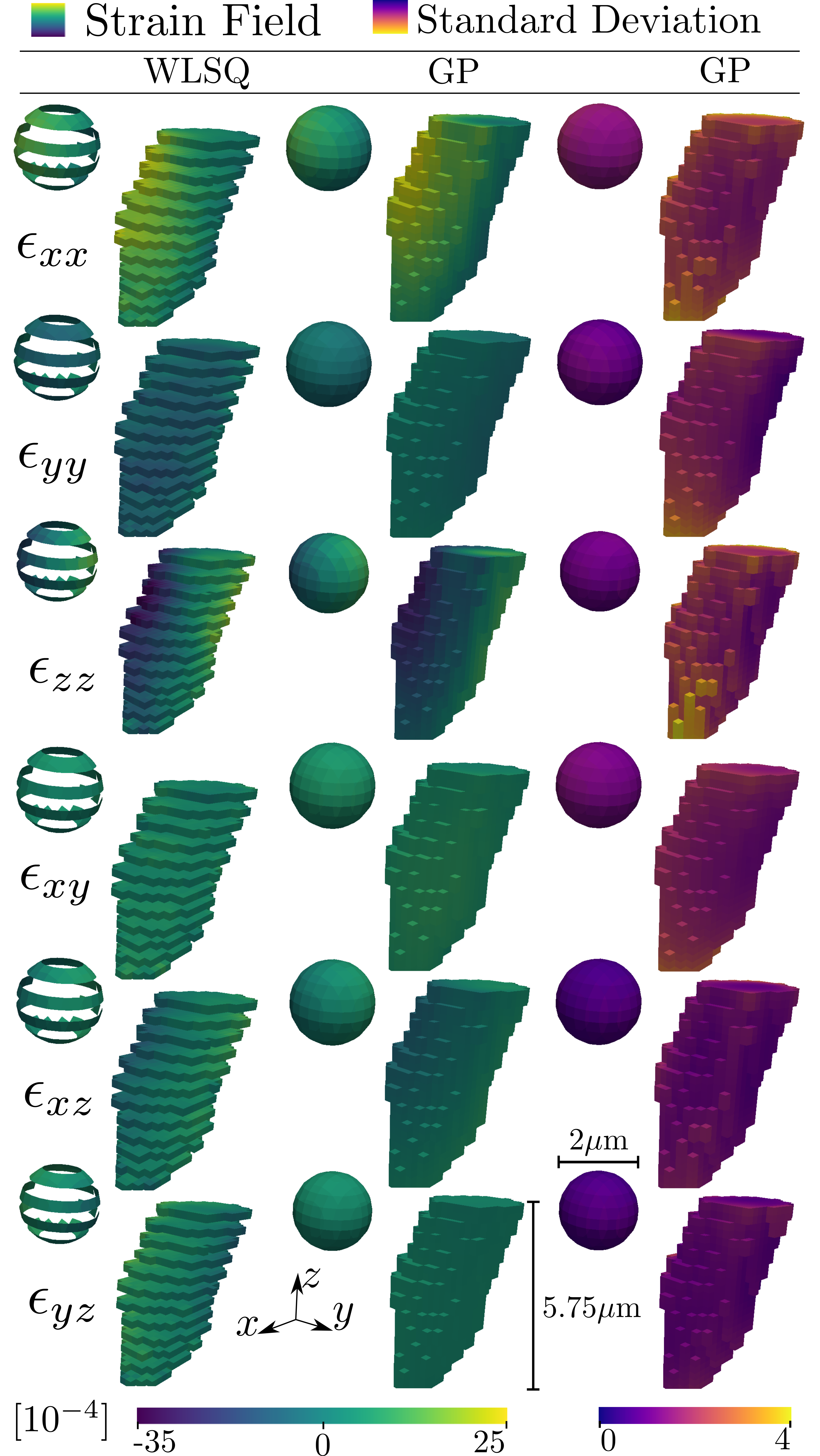}
    \caption{Reconstructed strain field using WLSQ (left column) and the GP method (middle column) of columnar tin grain embedded within a polycrystalline sample. The rightmost column depicts the estimated uncertainty of the GP reconstruction. The 3D surface of the voxelated grain is presented together with a pull out enlarged interior spherical cut with centre at the grain centroid and a radius of 1\(\mu\)m. Two separate colormaps have been assigned to enhance contrast for the various fields, however, units of strain remain the same across plots [x10\(^{-4}\)].}
    \label{fig:columnar_tin_grain}
\end{figure}

To further compare the GP and WLSQ reconstruction methods, analysis of a previously studied columnar tin grain has been included. This additional analysis further serves to show that the presented method is computationally feasible for state-of-the-art scanning-3DXRD data sets.
Including hyperparameter optimisation, the GP reconstruction was performed on a single CPU in a few minutes. The data for this example from \citep{Hektor2019} and the input experimental parameters are identical to those presented in Table \ref{tab:simul_params} except for the beam size, which was 0.25\(\mu\)m. Similarly, the relaxed lattice state was as defined in Table \ref{tab:lattice}.
In the original experiment, the X-ray beam was scanned across the \(x\)-\(y\) plane producing a space-filling map of measurements, however, due to time constraints, the data were collected for every second \(z\)-layer as seen in the rightmost column of Figure \ref{fig:columnar_tin_grain}. The reader is referred to the original publication \cite{Hektor2019} for further information on the experimental setup, sample, and preliminary data analysis. 

As the GP method uses a non local basis representation of the strain field \eqref{eq:fourier_basis} interpolation between measured slices is an automatic feature of the method. For the WLSQ method, although some interpolation scheme could be selected, we have selected to present the raw reconstructions, this also highlights the added benefit of the selected basis for the GP method. Hyper parameter optimisation and smoothness constraints for the WLSQ method were performed and selected as in section \ref{subsec:Single Crystal Simulation Test-case}.

\section{Discussion}\label{sec:Discussion}
Comparison of true and predicted fields in Figure \ref{fig:maxwell_simulation} for the two methods indicates that the reconstructions captured well the simulated input strain state.
For all strain components in table \ref{tab:rmse_and_mae}, both the RMSE and MAE are in the order of the expected experimentally-limited strain resolution (\(10^{-4}\)). We note, however, that the GP has consistently lower RMSE, MAE as well as maximum absolute errors in comparison to the WLSQ. The enhanced performance is believed to be attributed to the joint effect of the equilibrium prior, optimised correlation kernel and nonlocal basis selection.

The results of table \ref{tab:rmse_and_mae} indicate that, in general, the strain tensor \(z\)-components enjoy more accurate reconstructions compared to the \(xy\)-components. This observation is in line with previous work \citep{Margulies2002}; \citep{Lionheart2015}; \citep{Henningsson2020} and is
explained by the nonuniform sampling of strain taking place in scanning-3DXRD. The GP regression quantifies this phenomenon via the reconstructed standard deviation fields (Figure \ref{fig:maxwell_simulation} bottom row). Indeed the uncertainty in the predicted mean is elevated for the \(xx\)- and \(yy\)-components and show similar patterns as the residual fields. 

On the performance of the two methods, Figure \ref{fig:convergence_study_rmse_and_mae} indicates that fewer measurements are needed for the GP compared to the WLSQ whilst achieving a more accurate result. Little reduction in the RMSE and MAE is seen for the GP after about 50\% of the measurements have been introduced. This could imply that it is possible to retrieve approximations to the full strain tensor field from reduced scanning-3DXRD data sets. This could be attractive as scanning-3DXRD typically has time consuming measurement sequences.

It is evident that the reconstructed fields have maximum uncertainties at the boundary of the grain, as can be seen from the cutout spheres of Figure \ref{fig:maxwell_simulation} and \ref{fig:columnar_tin_grain}. The elevated standard deviation at the grain surface is explained by the tomographic measurement procedure, which has an increasing measurement density towards the grain centroid. Furthermore, as measurements do not exist outside of the grain, points lying on the grain surface will, in some sense, have a reduced number of points that they are correlated with.

The predicted strain field of the columnar tin grain of Figure \ref{fig:columnar_tin_grain} shows similar patterns between the two regression methods. The uncertainty is again seen to be reduced on the interior of the grain and the posterior standard deviation is in the order of the experimental strain resolution of \(10^{-4}\). This validates the applicability of GP regression on real state-of-the-art scanning-3DXRD synchrotron data.

\subsection{Outlook}\label{subsec:Outlook}
Two future potential improvements to strain predictions should be mentioned. Firstly, the selection of covariance function, although restricted to give a positive definite covariance matrix, is not unique; other selections may outperform the squared exponential kernel used here. Secondly, for polycrystalline samples, additional prior knowledge of grain boundary strain could be extracted by considering the total sample grain map and consider that tractions must cancel on the interfaces (i.e. incorporating and extending the work in \citep{hendriks2019implementation}). Two challenges with this exist (I) the uncertainty in reconstructed grain shapes leading to uncertainty in the interface normal and (II) uncertainty in the per-point grain orientation leading to uncertainty in the grain compliance. 

\section{Conclusions}\label{sec:Conclusions}
Intragranular strain estimation from scanning-3DXRD data using a Gaussian Process is shown to provide a new and effective strain reconstruction method. By selecting a continuous differentiable Fourier basis for the Beltrami stress functions, a static equilibrium prior can be incorporated into the reconstruction, guaranteeing that the predicted strain field will satisfy the balance of both angular and linear momentum. The regression procedure results in a per-point estimated mean strain as well as per-point standard deviations, providing new means of estimating the per-point uncertainty of the reconstruction. Furthermore, the proposed method incorporates the spatial structure of the strain field by making use of a generic covariance function, optimised by maximising the out-of-sample log likelihood. With the introduction of these three features, the equilibrium prior, the per-point uncertainty quantification and the optimised spatial smoothness constraints, the proposed regression method address weaknesses discussed in previously proposed reconstruction methods. Specifically, in comparison to a previously proposed weighted least squares approach, it is found, from numerical simulations, that the Gaussian Process regression consistently produces reconstructions with lower root mean squared errors, mean absolute errors and maximum absolute errors across strain components. Moreover, it is shown that the reconstruction error as a function of the number of available measurements is reduced for the Gaussian Process.

\textbf{Acknowledgements}
The authors are grateful for the beamtime
provided by the ESRF, beamline ID11, where the diffraction data were collected \cite{Hektor2019}. The authors would also like to thank Stephen Hall for valuable input on the manuscript.

\bibliography{bibliography}

\end{document}